\documentclass[aps,prd,floatfix,nofootinbib,superscriptaddress,twocolumn,showpacs,showkeys]{revtex4-2}

\usepackage[english]{babel}
\usepackage[utf8]{inputenc}       
\usepackage{mathrsfs}
\usepackage{flushend}
\usepackage{graphicx}
\usepackage{caption}
\usepackage{subcaption} 
\usepackage{float}  
\usepackage{xcolor}
\usepackage[export]{adjustbox}
\usepackage{wrapfig}
\graphicspath{{./images/}}
\usepackage{graphicx,epstopdf} 
\usepackage{amsmath,amsfonts,extarrows,amssymb,mathrsfs,times,bm}
\usepackage{extarrows}
\usepackage{enumitem}
\usepackage{comment}
\usepackage{mathtools}
\usepackage{siunitx}
\usepackage{wasysym}
\usepackage[colorlinks=true,breaklinks=true,linkcolor=blue,citecolor=blue,urlcolor=blue]{hyperref}   
\usepackage{tikz}
\usepackage[usenames,dvipsnames]{pstricks}
\usepackage{epsfig}
\usepackage{allpurpose}
\usepackage[utf8]{inputenc}     

\newcommand\nn{{\nonumber}}

\newcommand{\hatq}{\hat{q}}
\usepackage{ulem}  
\usepackage{url}  
\usepackage[thicklines]{cancel}   
\usepackage{color}

\newcommand{\delred}[1]{{\color{red}{\ifmmode\text{\sout{\ensuremath{#1}}}\else\sout{#1}\fi}}}

\begin{document}
\title{Periapsis shift in magnetized stationary and axisymmetric spacetimes }

\author{Yuhan Zhou}
\affiliation{School of Physics and Technology, Wuhan University, Wuhan, 430072, China}

\author{Junji Jia}
\email[Corresponding author:~]{junjijia@whu.edu.cn}
\affiliation{Department of Astronomy \& MOE Key Laboratory of Artificial Micro- and Nano-structures, School of Physics and Technology, Wuhan University, Wuhan, 430072, China}

\date{\today}

\begin{abstract}
In this work, we conduct a detailed study of the precession of charged particles in stationary and asymmetric spacetimes with external magnetic fields. Specifically, we develop the post-Newtonian method and the quasi-circular approximation to derive the periapsis shift respectively for two common types of magnetic fields, the dipolar one and the asymptotically uniform one. It is found using the PN method for magnetic fields decaying as fast or faster than a magnetic dipole that the magnetic effect in the periapsis shift appears from the same order of the traditional frame-dragging term due to the spacetime spin. The magnetic field is found to enhance (or decrease) the periapsis shift when the Lorentz force is attractive (or repulsive). When the repulsive Lorentz force is strong enough, the periapsis shift can become negative. For magnetic fields with a slower decay rate, the periapsis shift of quasi-circular orbits exhibits a more complex dependence on the Lorentz force. The periapsis shift increases as the attractive Lorentz force increases from zero but will decrease eventually. However, when the Lorentz force is repulsive, the orbit develops local helical loops and the periapsis shift approaches $-2\pi$. The results for the dipolar magnetic field are applied to the periapsis shift of Mercury around the Sun and the S2 around the Sgr A* to constrain the dipole moment of the center and the charge of the orbiting object. 
\end{abstract}

\keywords{periapsis shift, perihelion precession, post-Newtonian method, magnetic field, Lorentz force}
\maketitle

\section{Introduction} 

The successful explanation of the periapsis shift (PS) of Mercury by Einstein serves as a validation of General Relativity (GR) \cite{Einstein:1916}, which remains the most successful theory of gravitation to date. Subsequent astronomical observations have indicated that numerous other celestial bodies also exhibit PS, including all planets and satellites in the solar system \cite{Lucchesi_2010,Everitt_2011,Iorio_2019}, binary star systems (such as low-mass X-ray binaries) \cite{Damour_1988}, as well as stars like S2 orbiting the supermassive compact object Sagittarius A* (Sgr A*) at the center of our galaxy  \cite{GRAVITY:2020gka}. Research on PS of test particles around central bodies in simple spacetimes with only gravitational interaction has been carried out extensively, with studies conducted in Schwarzschild spacetime  \cite{Bini_2005,Vogt_2008,Schmidt_2011,Shchigolev_2016,Mak_2018,Walters_2018,Cao_2017}, Reissner-Nordström spacetime  \cite{Shchigolev_2015,Avalos_2012,Yapeng_2013,Michael_2022}, Kerr spacetime  \cite{Brill_1999,Colistete_2002,Bini_2005,Vogt_2008}, and Kerr-Newman spacetime  \cite{jiang_post-newtonian_2014,Heydari_Fard_2019} being well-known. However, recent experimental observations have revealed the presence of significant magnetic fields in the vicinity of black holes (BHs)  \cite{Dovciak_2004,Zaja_ek_2019,Daly_2019,YouBei_2023}. For instance, a strong magnetic field has been detected near the center of Sgr A*  \cite{eatough_strong_2013}. Additionally, the polarized image of the M87 BH released by the Event Horizon Telescope collaboration further supports this finding  \cite{Himwich_2020,EHT_M87_Telescope2,EHT_M87_Telescope3,EHT_SgrA_Telescope2,narayan_polarized_2021}. Although the mechanisms underlying the generation of external magnetic fields remain somewhat unclear, preliminary theoretical deductions suggest that they may include primordial magnetic fields originating from collapsed progenitor stars \cite{Ferrario_2015}, dynamo effects generated by accretion disks composed of substantial plasma surrounding BHs  \cite{dynamo_effects_2013,Latif_2016}, as well as background magnetic fields produced during the early stages of rapid cosmic expansion  \cite{Dario_2001,Jain_2012,Subramanian_2016}.

Due to the possibility that the test particles can carry very large specific charges, even very weak magnetic fields can have a non-negligible effect on the motion of charged particles, particularly on their PS. Previous studies examining the influence of electromagnetic interactions on PS have primarily considered the intrinsic charge of the central body \cite{Qianchuan_2023}. In contrast, the effect of external magnetic fields of the massive central bodies on the PS of charged test particles was not well studied, which becomes the main purpose of this work. 

In this study, we systematically classify magnetized stationary and axisymmetric (SAS) spacetimes, focusing on two common configurations: SAS spacetime with a dipolar magnetic field and the asymptotically uniform magnetic field, respectively. We employ two approaches to calculate the PS of test particles in each of them. One is the widely used and well-validated post-Newtonian (PN) expansion method, which assumes sufficiently large orbits, making it suitable for weak field limits. The other method is a quasi-circular method, based on the assumption of stable circular orbits, approximating the motion of particles to be closely near the circle, thereby transforming the problem of calculating the PS into solving for the orbital angular frequency and radial epicyclic frequency. 

The article is organized as follows. In Sec. \ref{sec:Dynamics}, we derive the equations of motion of charged particles in SAS spacetimes along with the four-potential of the axisymmetric electromagnetic field in general form, and define their Hamiltonian, effective potential and PS. In Sec. \ref{sec:PN method}, we developed the general PN method for the PS, whose results are applied to the Kerr BH with dipole magnetic field. Sec. \ref{sec:QPO} is dedicated to the quasi-circular method as well as its application in the uniformly magnetized Kerr BH case. In Sec. \ref{sec:application}, we apply the results in previous sections to the PS of Mercury around the Sun and S2 around Sgr A* to constrain their charges and effective dipolar moment. 
Finally, Sec. \ref{sec:conclusion} concludes the work. Throughout the paper, we adopt the spacetime signature convention $(-, +, +, +)$ and utilize the geometric units $G = c = 4 \pi \epsilon_0 = 1$. 

\section{Motion of charged particles} \label{sec:Dynamics}    
In this section, we study the motion of a charged particle in a general SAS spacetime described by the metric of the form
\begin{equation}
    \dd s^2 = -A \dd t^2 + B \dd r^2 + C \dd {\phi}^2 + 2D \dd t \dd\phi + G \dd {\theta}^2, \label{metric}
\end{equation}
where the functions $A,\, B, \,C, \,D, \,G$ are functions of $r$ and $\theta$ only. For the electromagnetic field in this spacetime, we limit it to only axisymmetric electromagnetic field described by electromagnetic 4-potential $(\mathcal{A}_t(r,\theta),\,0,\,0,\,\mathcal{A}_\phi(r,\theta))$. In this electromagnetic field, to confine the motion of charged particles to the equatorial plane, we will have to require that the only nonzero component is $B_{\hat{\theta}}$, perpendicular to the equatorial plane. 

In spacetime with an electromagnetic field, the motion of charged particles follows the Lorentz equation
\begin{equation}
    \frac{\dd U^{\mu}}{\dd{\tau}} + \Gamma ^{\mu }{}_{\nu \sigma }U^{\nu}U^{\sigma} =\frac{q}{m}F^{\mu}{}_{\nu}U^{\nu}, \label{Lorentz equation}
\end{equation}
where $U^\mu=\dd x^{\mu}/\dd\tau$ is the 4-velocity of the test particle with the normalization condition $U^\mu U_\mu = -1$, and $F_{\mu \nu } = \mathcal{A}_{\nu;\mu}-\mathcal{A}_{\mu;\nu}$ is the electromagnetic field tensor. $q$ and $m$ are the charge and mass of the test particles respectively. 

Here and in what follows, we will concentrate our study on the equatorial plane $\theta= \pi/2$. By substituting the metric Eq. \eqref{metric} and this condition into Eq. \eqref{Lorentz equation} and doing a first integration, we can easily derive the equations of motion in the equatorial plane
\begin{subequations}
\label{eq:tprdot}
    \begin{align}
    \dot{t}=&\frac{D\left(L-\hat{q} \mathcal{A}_{\phi }\right)+C(E+\hat{q} \mathcal{A}_t)}{D^2+AC}, \label{tdot}\\ 
    \dot{\phi}=&\frac{A\left(L-\hat{q} \mathcal{A}_{\phi }\right)-D(E+\hat{q} \mathcal{A}_t)}{D^2+AC}, \label{phidot}\\
    \dot{r}^2=&\frac{1}{B\left(D^2+AC\right)}\Big[C(E+\hat{q} \mathcal{A}_t)^2-A\left(L-\hat{q} \mathcal{A}_{\phi }\right)^2\notag\\
    &+2D(E+\hat{q} \mathcal{A}_t)\left(L-\hat{q} \mathcal{A}_{\phi }\right)\Big]-\frac{\kappa}{B}, \label{rdot2}
    \end{align}
\end{subequations}
with $E$ being the specific energy and $L$ being the particle's generalized specific angular momentum, while $\hat{q}\equiv q/m$ represents its specific charge. Parameter $\kappa$ takes the value of 1 because we study the charged particles here. Since we have specified $\theta=\pi/2$, the metric and 4-potential functions here and henceforth are functions of $r$ only. 

For the sake of the subsequent analysis, it is necessary to introduce the Hamiltonian of the charged particle  
\begin{equation}
    H=\frac{1}{2}g^{\alpha\beta}p_{\alpha}p_{\beta}+\frac{1}{2}m^2,   \label{Hamiltonian form 1}
\end{equation}
where $p^{\alpha}=m U^{\alpha}$ is the kinematical 4-momentum.
Thus, the Hamiltonian can be rewritten in the following form
\begin{equation}
    H=\frac{1}{2}g^{rr}p_{r}^2+V_{\mathrm{eff}},  \label{Hamiltonian form 2}
\end{equation}
with the last term being introduced as the potential part of the Hamiltonian. By subtracting its first term on the right side of Eq. \eqref{Hamiltonian form 1} and replacing the metric components $g^{\alpha\beta}$ in the expression with the metric functions from Eq. \eqref{metric} as well as $p_{\alpha}$ with Eq. \eqref{eq:tprdot}, we obtain the effective potential in the following form

\begin{align}
     V_{\mathrm{eff}} =& \frac{1}{2}\bigg\{1-\frac{1}{D^2+AC}\Big[C (E+\hat{q} \mathcal{A}_t)^2-A (L-\hat{q} \mathcal{A}_{\phi})^2\notag\\&+2D (E+\hat{q} \mathcal{A}_t)(L-\hat{q} \mathcal{A}_{\phi})\Big]\bigg\}. \label{Veff_SAS}
\end{align}

\section{PN method for the PS}   \label{sec:PN method}
\subsection{PS in general SAS spacetimes with magnetic field}

The PN method is an extensively used method for solving the PS  \cite{Gong_2009}. In this section, we try to solve the PS of charged particles in the SAS spacetimes with certain axisymmetric electromagnetic fields. The first assumption of the PN method is that the orbits of the particles can be regarded as some kind of perturbation of the Keplerian elliptical orbits
\begin{equation}
    u=\frac{1}{r}=\frac{1+e\cos \psi(\phi)}{p}, \label{eq:Elliptical orbit hypothesis}
\end{equation}
where $e$ is the eccentricity, $p$ is the semi-latus rectum, and $\psi(\phi)$ describes the deviation of orbits from the ellipse, of which the first-order approximation is exactly $\phi$. Obviously, this orbit has a minimum radius $r_1$, corresponding to $\psi=0$, and a maximum radius $r_2$ corresponding to $\psi=\pi$, i.e., 
\begin{align}
r_1=\frac{1-e}{p},\, r_2=\frac{1+e}{p}.   \label{eq:r1 r2}
\end{align}
It is worth noting that, as complicated as it may be, such radii of the perihelion and aphelion can always be obtained numerically from Eq. \eqref{rdot2} by setting 
\begin{align}
\dot{r}|_{r=r_{1,2}}=0. \label{eq:bcinr}
\end{align}
However, for general parameters $(\hat{q},\,E,\,L)$ with arbitrary metric functions and 4-potential, this equation may not admit closed-form solutions for  $r_{1,2}$.

From the equation of motion Eq. \eqref{eq:tprdot}, we can easily derive that 
\begin{align}
    \left( \frac{\dd u}{\dd \phi} \right) ^2&=u^4\frac{D^2+AC}{B[A(L-\hat{q} \mathcal{A}_{\phi})-D(E+\hat{q} \mathcal{A}_t)]^2}\notag\\
    &\times\Big[C(E+\hat{q} \mathcal{A}_t)^2-A(L-\hat{q} \mathcal{A}_{\phi})^2-(D^2+AC)\notag\\&+2D(E+\hat{q} \mathcal{A}_t)(L-\hat{q} \mathcal{A}_{\phi})\Big] \equiv F(u),    \label{eq:def F(u)}
\end{align}
where we have defined the right-hand side as a function of $u$.

The PS is defined as 
\begin{align}
\alpha_{\mathrm{PN}}=&
\int_{r_1}^{r_2} \dd \phi-2 \pi,  \label{eq:PN PS integral expression1}
\end{align}
and can be transformed to
\begin{align}
\alpha_{\mathrm{PN}}
=2\int_{0}^{\pi} \frac{\dd \phi}{\dd \psi} \dd\psi-2 \pi, \label{eq:PN PS integral expression2}
\end{align}
where the integrand after using Eqs. \eqref{eq:Elliptical orbit hypothesis} and \eqref{eq:def F(u)} becomes
\begin{equation}
    \frac{\dd \phi}{\dd \psi}=\frac{\dd \phi}{\dd u}\frac{\dd u}{\dd\psi}=-\frac{e}{p}\frac{\sin{\psi}}{\dd u/\dd \phi}=\frac{e}{p}\frac{\sin{\psi}}{\sqrt{F(u)}}.\label{eq:dphi/dpsi}
\end{equation}
The PS is then transformed to
\begin{align}
\alpha_{\mathrm{PN}}
=2\int_{0}^{\pi} \frac{e}{p}\frac{\sin{\psi}}{\sqrt{F(u)}} \dd\psi-2 \pi. \label{eq:PN PS integral expression3}
\end{align}

Since $F(u)$ as defined in Eq. \eqref{eq:def F(u)} is quite complicated, this integral usually can not be carried out to obtain a closed form result. In the PN limit however,
the smallness of $u$ enables us to expand $F(u)$ as a power series of $u$
\begin{equation}
    F\left( u \right) =\sum_{n=0}^{\infty}{f_n u^n}. \label{eq:F(u) in series form}
\end{equation}
The expansion coefficients are naturally linked to the metric function and potential functions $\mathcal{A}_t,\, \mathcal{A}_\phi$. Assuming that they have asymptotic expansions of the form
\begin{align}
    A(r)&=\sum_{n=0}^{\infty}{\frac{a_n}{r^n}},\  B(r)=\sum_{n=0}^{\infty}{\frac{b_n}{r^n}},\notag\\
    C(r)&=\sum_{n=0}^{\infty}{\frac{c_n}{r^{n-2}}},\  D(r)=\sum_{n=1}^{\infty}{\frac{d_n}{r^n}},\notag\\
    \mathcal{A}_t(r)&=\sum_{n=1}^{\infty}{\frac{q_n}{r^{n}}},\  \mathcal{A}_\phi(r)=\sum_{n=0}^{\infty}{\frac{\mu_n}{r^{n}}}, \label{series expansion of metric function}
\end{align}
with the additional condition $a_0=b_0=c_0=1$ due to the asymptotical flatness of the spacetimes we focus on. The $q_1$ here plays the role of the total charge. We can also set $\mu_0=0$ because $\mathcal{A}_\phi$ always appears with $L$ in the motion equations Eq. \eqref{eq:tprdot} and its constant can be absorbed into $L$. Even if $\mu_0$ is kept, it can be verified that generally the final expressions for the PS will not depend on it. The coefficient $\mu_1$ can actually be identified as the magnetic dipole moment because the corresponding $B_{\hat{\theta}}$ component of the magnetic field on the equatorial plane is
\begin{align}
B_{\hat{\theta}}=\frac{F_{r\phi }}{r\sin\theta}\approx \frac{\mu_1}{r^3}+\mathcal{O}(r^{-4}).
\end{align}
Then we can derive the coefficients $f_n$ as functions of $a_n,\, b_n,\, c_n,\, \mu_n$ as well as $E,\, L,\, \hat{q}$. The first few orders of coefficients $f_n$ are
\begin{subequations}
\label{eq:fn3}
    \begin{align}
    f_0 =& \frac{E^2-1}{L^2}\label{f0a},\\
    f_1 =& \frac{2E \hat{q}q_1-a_1}{L^2}+\left(E^2-1\right)\left[\frac{a_1-b_1+2c_1}{L^2}\right.\notag\\
    &\left.-\frac{2 \left(a_1 L-d_1 E-\hat{q}\mu _1\right)}{L^3}\right] .\label{f1b}
    \end{align}   \label{eq:fn}
\end{subequations}

We next continue to simplify $F(u)$ in Eq. \eqref{eq:F(u) in series form} in order to eventually carry out the integral \eqref{eq:PN PS integral expression3}. Substituting $u$ in Eq. \eqref{eq:Elliptical orbit hypothesis} into right-hand side of Eq. \eqref{eq:F(u) in series form}, it becomes
\begin{equation}
F(u)=\sum_{n=0}^{\infty}{\frac{f_n}{p^n}}\left( 1+e\cos \psi \right) ^n. \label{F(u) in substituted series form}
\end{equation}
Using the following identities of integer powers of $\cos{\psi}$
\begin{subequations}
    \begin{align}
    \cos ^{2m}\psi &=1-\sin ^2\psi \sum_{k=1}^m{C_{m}^{k}\left( \cos ^2\psi -1 \right) ^{k-1}},\nn\\ 
    \cos ^{2m+1}\psi &=\cos \psi-\sin ^2\psi \sum_{k=1}^m{C_{m}^{k}\left( \cos ^2\psi -1 \right) ^{k-1}\cos \psi},\nn
    \end{align}
\end{subequations}
we recognize that $F(u)$ can always be recast into the following form
\begin{align}
    F(u)=&F_1(E,\ L,\ e,\ p)+F_c(E,\ L,\ e,\ p)\cos{\psi}\notag\\
    &+F_s(E,\ L,\ e,\ p,\ \cos{\psi})\sin^2{\psi}, \label{eq:F(u）=F1+Fc+Fs}
\end{align}
where $F_1,\, F_c,\, F_s$ denotes terms in $F(u)$ that are independent of $\psi$, linear to $\cos\psi$ and linear to $\sin^2\psi$.  Clearly these series functions are not difficult to obtain and should be linear combinations of $f_n$ with coefficients being polynomials of $e$ and $1/p$. In particular, $F_s$ to the first few orders of $1/p$ is found to take the form 
\begin{align}
    F_s=&-\frac{e^2 f_2}{p^2}-\frac{(e^3 \cos\psi+3e^2)f_3}{p^3}\notag\\
    &-\frac{(e^4 \cos^2\psi+e^4+4 e^3 \cos \psi+6 e^2)f_4}{p^4}+\mathcal{O}(p^{-5}).  \label{eq:fsfirst}  
\end{align}
In the meanwhile, utilizing the boundary condition \eqref{eq:bcinr} and definition \eqref{eq:Elliptical orbit hypothesis}, \eqref{eq:def F(u)} of $F(u)$, we find
\begin{equation}
    F(u)|_{\psi=0}=F(u)|_{\psi=\pi}=0 .  \label{eq:boundary condition}
\end{equation}
Applying this condition to Eq. \eqref{eq:F(u）=F1+Fc+Fs} immediately produces 
\begin{align}
\label{eq:f1fc0}    F_1(E,\ L,\ e,\ p)=F_c(E,\ L,\ e,\ p)=0,
\end{align}
and the $F(u)$ finally simplifies to
\begin{align}
F(u)=F_s(E,\ L,\ e,\ p,\ \cos{\psi})\sin^2{\psi}.\label{fuin}
\end{align}

Substituting Eq. \eqref{fuin} and the series \eqref{eq:fsfirst} for $F_s$ into the PS formula \eqref{eq:PN PS integral expression3}, the integrand can be further expanded in the large $p$ limit as    
\begin{align}
    \frac{e}{p}\frac{\sin\psi}{\sqrt{F_s}}=&\frac{1}{\sqrt{- f_2}}+\frac{f_3 (e \cos \psi +3)}{2 \left(-f_2\right){}^{3/2}}\frac{1}{p} + \frac{1}{8 \left(-f_2\right)^{5/2}p^2}\notag\\
    &\times\Big[3 f_3^2 (e \cos \psi +3)^2-2 f_2 f_4 \left(e^2 \cos 2 \psi +3 e^2 \right.\notag\\
    &\left.+8 e \cos \psi +12\right) \Big] +\mathcal{O}(p^{-3}).  \label{eq:series d phi/d psi}
\end{align}
Carrying out the $\psi$ integral which is simple now, one finds the following result for the PS as

\begin{align}
    \alpha_{\mathrm{PN}}&=
    -2 \pi+\frac{2\pi}{\sqrt{-f_2}}+\frac{3 \pi f_3}{\left(-f_2\right){}^{3/2} p}\notag\\
    &+ \frac{3 \pi  \left[\left(e^2+18\right) f_3{}^2-4 \left(e^2+4\right) f_2 f_4\right]}{8 \left(-f_2\right){}^{5/2} p^2}\notag\\
    &-\frac{3 \pi  \left[4 \left(13 e^2+36\right) f_2 f_3 f_4-15 \left(e^2+6\right) f_3{}^3\right]}{16 \left(-f_2\right){}^{7/2} p^3}\notag\\
    &+ \mathcal{O}\left(\frac{1}{p}\right)^4. \label{eq:pspro}
\end{align}

In this PS however, the functions $f_n$ as given in Eq. \eqref{eq:fn} are functions of not only kinetic variables $(e,\,p)$, but also $(E,\,L)$. However, as stated previously, these two sets of kinetic parameters are equivalent and therefore redundant in Eq. \eqref{eq:pspro}. Therefore, next we work out the expression of $(E,\,L)$ in terms of $(e,\,p)$ so that the final PS can be expressed completely in terms of the latter.

The explicit forms of $E$ and $L$ in terms of $e$ and $p$ in the most general case are not easy to obtain. However, since we are working in a PN approximation, it is always possible to find the asymptotic form for $(E, L)$ as power series of $1/p$. To be more concrete, we assume the following series form 
\begin{equation}
    E=\sum_{n=0}^{\infty}{\frac{e_n}{p^{n/2}}},\ L=\sum_{n=-1}^{\infty}{\frac{l_n}{p^{n/2}}}, \label{eq:EL series primary form}
\end{equation}
whose coefficients can be found using the undetermined coefficient method. 
Substituting them into Eq. \eqref{eq:f1fc0} matching the coefficients of each order of $p$ allow us to solve the coefficients $e_n$ and $l_n$ in Eq. \eqref{eq:EL series primary form} iteratively. We show the first few orders of them as 
\begin{subequations}
    \begin{align}
    E =& 1 - \frac{1}{4} \left(a_1-2q q_1\right) \left(e^2-1\right) \frac{1}{p}+ \frac{1}{32} \left(a_1-2q q_1\right)\notag\\
    &\times\left(e^2-1\right)^2 \left(3 a_1-4 c_1-2q q_1\right)\frac{1}{p^2}+ O\left(p\right)^{-5/2}, \\
    L =& \sqrt{(\frac{-a_1}{2}+ q q_1)p} + \frac{1}{4\sqrt{2p(-a_1+2q q_1)}}\notag\\
    &\times\bigg\{(e^2+3)a_1^2+2q\Big[(e^2+3)c_1 q_1+(e^2+1)q q_1^2+4q_2\Big]\notag\\&-4a_2-a_1\Big[(e^2+3)c_1+(3e^2+5)q q_1\Big]\bigg\}\notag\\&+ \frac{ \left(e^2+3\right) q \mu _2}{2p} + O\left(p\right)^{-3/2},  \label{eq:series E,L}
    \end{align}  
\end{subequations}
where without losing any generality, we have chosen a positive sign when taking square root for $L$, i.e., we define the rotation direction of the particle to be the $+\hat{z}$ direction. The PS value when the particle orbits the center in the opposite direction can be obtained by switching the magnetic field and spacetime spin directions while keeping the spacetime electric potential unchanged. 

Substituting Eq. \eqref{eq:series E,L} into \eqref{eq:pspro}, the final PS for charged particles in a general SAS spacetime with a axisymmetric magnetic field becomes

\begin{widetext}
\begin{align}
    \alpha_{\mathrm{PN}}=&\frac{\pi \left[a_1 b_1+a_1 c_1-2 a_1^2+2 a_2 + \hat{q} q_1\left(4a_1 -2b_1 -2c_1 \right)-2 \hat{q}^2 q_1^2-4 \hat{q} q_2\right]}{\left(a_1-2\hat{q} q_1\right)p}+\frac{4  \left[\sqrt{2} \pi  \left(d_1+ \hat{q}\mu _1\right)\right]}{\left(-a_1+2 \hat{q} q_1\right){}^{1/2}p^{3/2}}\notag\\
    &-\frac{\pi}{8 \left(a_1-2\hat{q} q_1\right){}^2 p^2}\bigg\{\Big[8 a_1(a_2 b_1+4 a_2 c_1 + 6 a_3)+2a_1^2(-40 a_2+4 b_2+16 c_2- b_1^2+2 b_1 c_1-2 c_1^2)\notag\\
    &-8 a_1^3 (b_1+4 c_1)+40 a_1^4-8 a_2^2+\hat{q}q_1\big(8a_1(31 a_2-4  b_2-16  c_2+  b_1^2+  c_1^2-2  b_1 c_1)+16(4 a_1^2- a_2)( b_1+4 c_1)\notag\\
    &-96 a_3-160 a_1^3\big)+\hat{q}q_2\big(32 a_2+72 a_1^2-16 a_1( b_1-4 c_1)\big)-96 a_1\hat{q}q_3+\hat{q}^2q_1^2(-160 a_2+32 b_2+128 c_2\notag\\
    &+200 a_1^2-8 b_1^2-8 c_1^2-40 a_1 b_1-160 a_1 c_1+16 b_1 c_1)+\hat{q}^2q_1q_2(-240 a_1+32 b_1+128 c_1)-32 \hat{q}^2q_2^2\notag\\
    &+192\hat{q}^2 q_1 q_3+\hat{q}^3 q_1^3(-88 a_1+16 b_1+64 c_1)+160 \hat{q}^3q_1^2 q_2+8 q^4 q_1^4\Big]+e^2 \Big[a_1^2(-2 b_1 c_1+4  b_2- b_1^2\notag\\
    &-5  c_1^2+4  c_2)+8 a_1^3 c_1-8 a_2 a_1 c_1+\hat{q}q_1 \big(4a_1(2 b_1 c_1 + b_1^2 -4 b_2+5 c_1^2 -4 c_2) -32 a_1^2 c_1  +16 a_2 c_1 +8 a_2 a_1 \big)\notag\\
    &+\hat{q}q_2(16 a_1 c_1-8 a_1^2)+\hat{q}^2 q_1^2(40 a_1 c_1 -8 a_1^2 -16 a_2 -8 b_1 c_1 -4 b_1^2 +16 b_2 -20 c_1^2 +16 c_2 ) +\hat{q}^2 q_1 q_2 \notag\\
    &\times(16 a_1 -32 c_1)+\hat{q}^3 q_1^3\left(24 a_1 -16 c_1 \right)-16 \hat{q}^4 q_1^4\Big]\bigg\}+\mathcal{O}(p)^{-5/2}.  \label{eq:PN PS general series expansion}
\end{align}
\end{widetext}
There are a few comments in order here. 
First is that this formula for the PS holds for charged particles in all SAS spacetimes with an electric field and an axisymmetric magnetic field decaying asymptotically at order $r^{-3}$ or faster as $r$ approaches infinity. The second is that to the leading order (order $p^{-1}$), the numerator can be split into the gravitational part and the electric contributions, i.e., the terms proportional to $\hat{q}$. Moreover, and somewhat surprising is that the electric interaction of $\hatq q_1$ modifies the entire leading order through its appearance in the denominator. The third point is that the magnetic contribution, in this case appears from half order of $p$ higher, i.e., the term proportional to $\hatq\mu_1/p^{3/2}$. In addition, the gravitation and electrostatic interaction also contribute to this order.  

\subsection{PS in Kerr spacetime with a dipole magnetic field}

In this subsection, we apply the method and results in the previous subsection to special configurations of a Kerr BH
magnetized by an external dipole magnetic field.  
The Kerr metric in the Boyer-Lindquist coordinates is given by \cite{Boyer_Lindquist_Metric}
\begin{align}
    \dd s^2 = &-\left(1 - \frac{2Mr}{\Sigma}\right)\dd t^2 - \frac{4aMr\sin^2\theta} 
    {\Sigma}\dd t \dd \phi + \frac{\Sigma}{\Delta}\dd r^2 \notag\\
    &+ \Sigma \dd\theta^2 + \left(\Delta + \frac{2Mr(r^2+a^2)}{\Sigma}\right)\sin^2\theta \dd\phi^2,  \label{Kerr metric}
\end{align}
where 
\begin{equation}
    \Sigma=r^2+a^2\cos^2\theta,\ \Delta=r^2-2Mr+a^2.
\end{equation}
and $a$ is the angular momentum per unit mass of the BH. 

For the four-potential of the electromagnetic field, 
a commonly used one is given by  \cite{Prasanna_1977,Prasanna_1978}
\begin{subequations}
\label{eq:kfp}
\begin{align}
    \mathcal{A}_{t}=&-\frac{3a\mu}{2\gamma^2\Sigma}\bigg\{\Big[r(r-M)+(a^2-Mr)\cos^2\theta\Big]\notag\\
    &\times\frac{1}{2\gamma}\ln{\left(\frac{r-M+\gamma}{r-M-\gamma}\right)}-(r-M\cos^2\theta)\bigg\}, \label{Kerr dipole 4-potential t component}\\
    \mathcal{A}_{\phi}=&-\frac{3\mu\sin^2{\theta}}{4\gamma^2\Sigma}\bigg\{(r-M)a^2\cos^2{\theta}+r(r^2+Mr\notag\\
    &+2a^2)-\Big[r(r^3-2M a^2+a^2r)+\Delta a^2\cos^2\theta\Big]\notag\\
    &\times\frac{1}{2\gamma}\ln{\left(\frac{r-M+\gamma}{r-M-\gamma}\right)}\bigg\}, \label{Kerr dipole 4-potential phi component}\\
    \mathcal{A}_r=&\mathcal{A}_{\phi}=0,
\end{align}
\end{subequations}
where $\gamma=\sqrt{M^2-a^2}$, and $\mu$ is the magnetic dipole moment. When $a$ is set to zero, the spacetime reduces to the Schwarzschild one and the electric component vanishes in the four potential \eqref{eq:kfp}, i.e., only the magnetic field survives.  The asymptotic expansion of the $\mathcal{A}_t$ and $\mathcal{A}_\phi$ components of the above four-potential on the equatorial plane takes the form
\begin{subequations}\label{eq:fpexp} 
    \begin{align}
    \mathcal{A}_t=&-\frac{a \mu }{2 r^3}-\frac{a \mu  M}{r^4}+\frac{3 a \mu  \left(a^2-6 M^2\right)}{10 r^5}+\mathcal{O}(r)^{-6},\\
    \mathcal{A}_{\phi}=&\frac{\mu }{r}+\frac{3 \mu  M}{2 r^2}+\frac{\mu  \left(a^2+24 M^2\right)}{10 r^3}-\frac{\mu  M \left(a^2-8 M^2\right)}{2 r^4}\notag\\&-\frac{3 \left(\mu  \left(a^4+54 a^2 M^2-160 M^4\right)\right)}{70 r^5}+\mathcal{O}(r)^{-6}.   \label{eq:aphiexp}
    \end{align}
\end{subequations}
It is clear that the electrostatic potential is always proportional to $a^i\mu$ where $i=1,3,\cdots$ while the magnetic one is proportional to $a^j \mu$ where $j=0,2,\cdots$. The contributions of these potentials in the PS, as seen in Eq. \eqref{eq:PS-Kerr-Dipole}, will be multiplied with the particle's specific charge $\hatq$ and the above features will help us to distinguish the gravitational, electrostatic and magnetic contributions from each other. In addition, let us also point out that the $\mathcal{A}_\phi$ in Eq. \eqref{eq:aphiexp} corresponds to an asymptotic electromagnetic field on the equatorial plane of the form
\begin{align}
    &E_{\hat{r}}\Big|_{\theta=\pi/2}\approx F_{rt}\Big|_{\theta=\pi/2}=\frac{3a\mu}{2r^4}+\mathcal{O}(r^{-5}),\\
    &B_{\hat{\theta}}\Big|_{\theta=\pi/2}\approx\frac{F_{\phi r}}{r\sin\theta}\bigg|_{\theta=\pi/2}=\frac{\mu}{r^3}+\mathcal{O}(r^{-4}) , \label{eq:magonplane}
\end{align}
a magnetic field along the $\hat{z}^-$ direction when $\mu$ is positive.

Expanding the metric functions in Eq. \eqref{Kerr metric} on the equatorial plane and into large $r$ series and substituting the expansion coefficients as well as those of Eq. \eqref{eq:fpexp} into Eq. \eqref{eq:PN PS general series expansion}, the PS in this case is then found as
\begin{widetext}
\begin{align}
    \alpha_{\mathrm{KD}}=&\frac{6 \pi  M}{p}+\frac{4 \pi s(\hat{q} \mu-2 a M)}{\sqrt{M}p^{3/2}}+\frac{3 \pi  \left[2 a^2 M-2 a \hat{q} \mu +\left(e^2+18\right) M^3\right]}{2 M p^2}-\frac{3\pi s\sqrt{M} \left[24 a M+\left(e^2-11\right) \hat{q} \mu \right]}{p^{5/2}}\notag\\
    &+\frac{3 \pi  }{2 M p^3}\Big[2 \left(25-3e^2\right) a^2 M^2+15 \left(e^2+6\right) M^4+\left(6 e^2-43\right)a \hat{q} \mu  M-2 \left(e^2-3\right) \hat{q}^2 \mu ^2 \Big]\notag\\
    &-\frac{3 \pi s  }{5 M^{3/2} p^{7/2}}\Big[-20 \left(e^2-3\right) a^3 M^2+90\left(e^2+ 10\right)a M^4 + \left(18 e^2-73\right) a^2 \hat{q} \mu M\notag\\&+4\left(3 e^2-98\right) \hat{q} \mu  M^3-5 \left(e^2-4\right)a \hat{q}^2 \mu ^2 \Big]+\mathcal{O}(p)^{-4}. \label{eq:PS-Kerr-Dipole}
\end{align}
\end{widetext}
where we have restored the dependence of the PS on the rotation direction of the particle using sign $s$, i.e., $s=1$ for counterclockwise rotation and $s=-1$ for clockwise rotation. 
As discussed after Eq. \eqref{eq:fpexp}, the terms proportional to $(a^i\hatq\mu)^n$ with $i=1,3,\cdots$ and $n=1,2,\cdots$ are contributions from electrostatic interaction while those proportional to  $(a^j\hatq\mu)^n$ with $j=0,2,\cdots$ and $n=1,2,\cdots$ are contributions from magnetic Lorentz forces.
Therefore, we immediately see from the above PS that the magnetic contribution starts to appear from the $p^{-3/2}$ order (the term containing $\hatq \mu$), which is half order higher than the PS due to electrostatic interaction between charged spacetime and particle \cite{Qianchuan_2023}. Moreover, since the magnetic field for positive $\mu$ is along the $\hat{z}^-$ direction, a positive and clockwise rotating charge will experience an inward Lorentz force and therefore effectively increase the attraction due to the gravitational. This explains the same signs between the leading $M/p$ term and the $\hatq\mu/(\sqrt{M}p^{3/2})$ term. 
Moreover, the magnetic contribution also appears in the $p^{-5/2},\,p^{-3}$ and $p^{-7/2}$ orders. The contributions from pure electrostatic interaction appear in the $p^{-2}$ order (term containing $q\hatq\mu$) and $p^{-3}$ order. Also interestingly, we observe that in the $p^{-7/2}$ order there is an electro-magnetic coupling term proportional to $a\hatq^2\mu^2$. 

By setting $\mu$ (or $\hat{q}$) to zero, we can eliminate the electromagnetic contribution and obtain the PS $\alpha_{\mathrm{K}}$ in Kerr spacetime 
\begin{widetext}
\begin{align}
    \alpha_{\mathrm{K}}=&\frac{6 \pi  M}{p}-\frac{8\pi s a \sqrt{M}}{p^{3/2}} +\frac{3 \pi  \left(2 a^2 +\left(e^2+18\right) M^2\right)}{2 p^2}-\frac{72\pi s a M^{3/2}}{p^{5/2}}+\frac{3 \pi}{2 p^3}\Big[2 \left(25-3 e^2\right)a^2 M+15 \left(e^2+6\right) M^3\Big]\notag\\&-\frac{6\pi s \sqrt{M}}{p^{7/2}}\Big[2\left(3-e^2\right) a^3 +9\left(10+ e^2\right)a M^2\Big] +\frac{3 \pi}{32 p^4}\Big[24 \left(3-2 e^2\right)a^4-20 \left(e^4+2 e^2-492\right) a^2  M^2 \notag\\&+35 \left(e^4+72 e^2+216\right) M^4\Big]+\mathcal{O}(p)^{-9/2}.  \label{eq:PS-Kerr}
\end{align}
\end{widetext}
The lower order formula for this result, to orders $p^{-3/2}$, $p^{-2}$ and $p^{-4}$, were found in Refs. \cite{Brill_1999}, \cite{Colistete_2002, Heydari_Fard_2019} and \cite{He_2024} respectively. Setting $a=0$ in Eq. \eqref{eq:PS-Kerr-Dipole}, we will obtain the PS $\alpha_{\mathrm{SD}}$ of a charge particle in a Schwarzschild spacetime with influence from a magnetic dipole field
\begin{widetext}
\begin{align}
    \alpha_{\mathrm{SD}}&=\frac{6 \pi  M}{p}+\frac{4 \pi s\hat{q} \mu}{\sqrt{M}p^{3/2}}+\frac{3 \pi  \left(e^2+18\right) M^2}{2 p^2}-\frac{3\pi s \left(e^2-11\right) \hat{q} \mu \sqrt{M}}{p^{5/2}}+\frac{3 \pi}{2 M p^3}\Big[15 \left(e^2+6\right) M^4-2 \left(e^2-3\right) \hat{q}^2 \mu ^2 \Big]\notag\\
    &+\frac{12 \pi s\left(3 e^2-98\right) \hat{q}\mu  M^{3/2}}{5 p^{7/2}}+\frac{3 \pi}{32 p^4}\Big[35 \left(e^4+72 e^2+216\right) M^4+16 \left(e^4-20 e^2+87\right)\hat{q}^2 \mu ^2 \Big]+\mathcal{O}(p)^{-9/2}. 
\label{PS-Schwarz-Dipole}
\end{align}
\end{widetext}

If we further eliminate the Lorentz force from Eq. \eqref{PS-Schwarz-Dipole} by setting $\hat{q}$ or $\mu$ to zero, the above result degenerates into the series expression for the PS of particles in pure Schwarzschild spacetime
\begin{equation}
\begin{split}
    \alpha_{\mathrm{S}}=&\frac{6 \pi  M}{p}+\frac{3 \pi \left(e^2+18\right) M^2}{2 p^2}+\frac{45 \pi \left(e^2+6\right) M^3 }{2 p^3}\\&
    +\frac{105 \pi \left(e^4+72 e^2+216\right) M^4}{32 p^4}+\mathcal{O}(p)^{-5},  \label{PS-Schwarz}
\end{split}
\end{equation}
whose first orders were computed in Refs. \cite{Shchigolev_2015,Shchigolev_2016,Mak_2018,Michael_2022}, and first three orders coincide with Eq. (36) in Ref. \cite{Heydari_Fard_2019} and the large $p$ expansion of Eq. (5) in Ref. \cite{Walters_2018}.

\begin{figure}[htp]  
    \centering
    \begin{subfigure}{0.5\textwidth}
        \centering
        \includegraphics[width=0.85\textwidth]{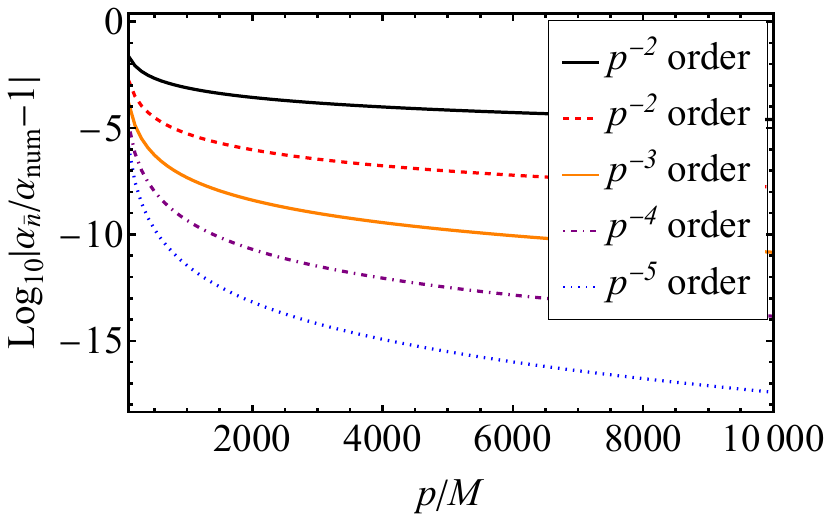}
        \caption{}
        \label{fig:PN-Error-p}
    \end{subfigure}
    \begin{subfigure}{0.5\textwidth}
        \centering
        \includegraphics[width=0.8\textwidth]{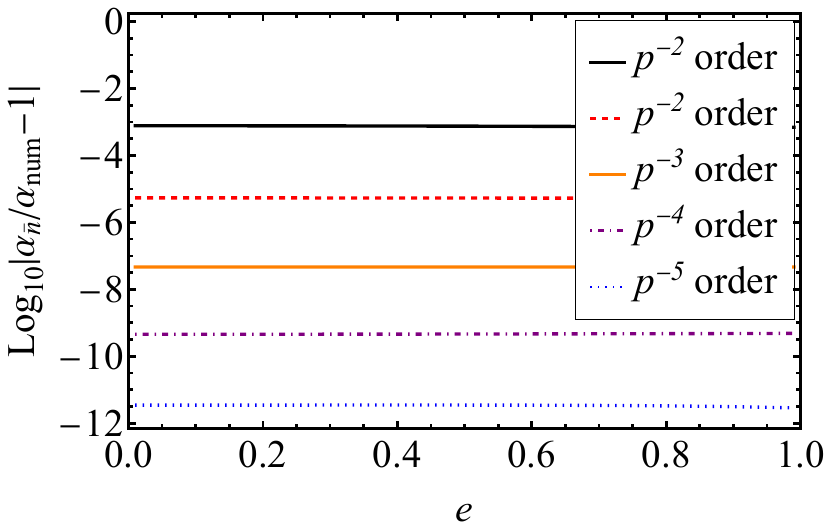}
        \caption{}
        \label{fig:PN-Error-e}
    \end{subfigure}
    \caption{Comparison of the numerical result with the PN method result Eq. \eqref{eq:PS-Kerr-Dipole} truncated to different orders. The default parameters are $e = 1/10$ (a), $p = 1000M$ (b) and $a = M/5$, $\hat{q} \mu=5$, $M = 1$, $s=1$.}
    \label{fig:PN-Error}
\end{figure}

To further examine the validity and accuracy of our PS result, namely, Eq. \eqref{eq:PS-Kerr-Dipole}, next we compute the PS using numerical integration from definition \eqref{eq:PN PS integral expression1} and compare with our PS formulas. 
Since the numerical integration was carried out to high accuracy, the corresponding $\alpha_{\mathrm{num}}$ can be considered as the true value of PS. In Fig. \ref{fig:PN-Error}, we illustrate the relative difference of our $\alpha_{\mathrm{KD}}$ in Eq. \eqref{eq:PS-Kerr-Dipole} when truncated to different orders with the numerical PS $\alpha_{\mathrm{num}}$, as functions of $p$ (Fig. \ref{fig:PN-Error} (a)) and $e$ (Fig. \ref{fig:PN-Error} (b)). It is seen from both plots that as the truncated order increases, the PN results converge to its true value exponentially. Fig. \ref{fig:PN-Error} (a) shows that as parameter $p$ increases, the precision of any PS of fixed truncation order improves in a power series form. However, the nearly horizontal lines in Fig. \ref{fig:PN-Error} (b) indicate that the precision is not particularly sensitive to eccentricity $e$. These features are understandable in that our result is a series expansion of parameter $M/p$ but no eccentricity approximation is used. 

After confirming the high accuracy of the PS results derived from the PN method, we can proceed to use it to study the effect of kinetic parameters $(p,\,e)$ of the motion and the intrinsic parameters of the spacetime and field $(a,\,\mu)$ and the particle ($\hatq$) on the PS. 
Generally speaking, the effect of $p$ is very similar to the Kerr case without electromagnetic field \cite{He_2024} and the KN case \cite{He_2024}, both qualitatively and quantitatively. Qualitatively, the PS in all these cases will decrease monotonically as $p$ increases because our PS after all is a power series in $p$. Quantitatively, the PS is largely determined by the first few leading orders, while the leading order coefficient is determined by the spacetime mass in the current case and the Kerr case without electromagnetic field \cite{He_2024}, as seen in the first term of Eqs. \eqref{eq:PS-Kerr-Dipole} and \eqref{eq:PS-Kerr}. Only in the case there is an electrostatic interaction between the spacetime and particle charges, the leading order coefficient might be changed slightly,  see Eq. (48) of Ref. \cite{Qianchuan_2023}. For the eccentricity $e$, from Eq. \eqref{eq:PS-Kerr-Dipole} we see that it only appears in the PS from the $p^{-2}$ order with a positive coefficient. Therefore, the PS only grows very weakly as $e$ increases when $p/M$ is large, which is typical in most applications.

The effect of parameters $(a,\,\mu,\,\hatq)$ on the PS is, in principle, more interesting. It is observed from Eq. \eqref{eq:PS-Kerr-Dipole} that the first contribution of $a$ is present in the next-to-leading order ($p^{-3/2}$ order) through the gravitational channel. This is the famous frame-dragging (FD) term, or Einstein-Lense-Thirring or gravitomagnetic term. Its effect has been studied in detail in the PS of neutral particles in ordinary Kerr or KN spacetimes  \cite{He_2024}. However, unlike these cases, the spin parameter now also appears in the electromagnetic potential \eqref{eq:fpexp} and therefore also contributes to the PS through these channels. The first non-gravitational contribution containing $a$, appearing in the $p^{-2}$ order, is actually an electrostatic contribution and the first magnetic contribution containing $a$ shows up in the $p^{-7/2}$ order (the term proportional to $a^2\hatq\mu$). Unfortunately, these orders are all higher than the gravitational one, and therefore we will not focus on their effect here. 

\begin{figure}[htp!]
    \centering
\includegraphics[width=0.45\textwidth]{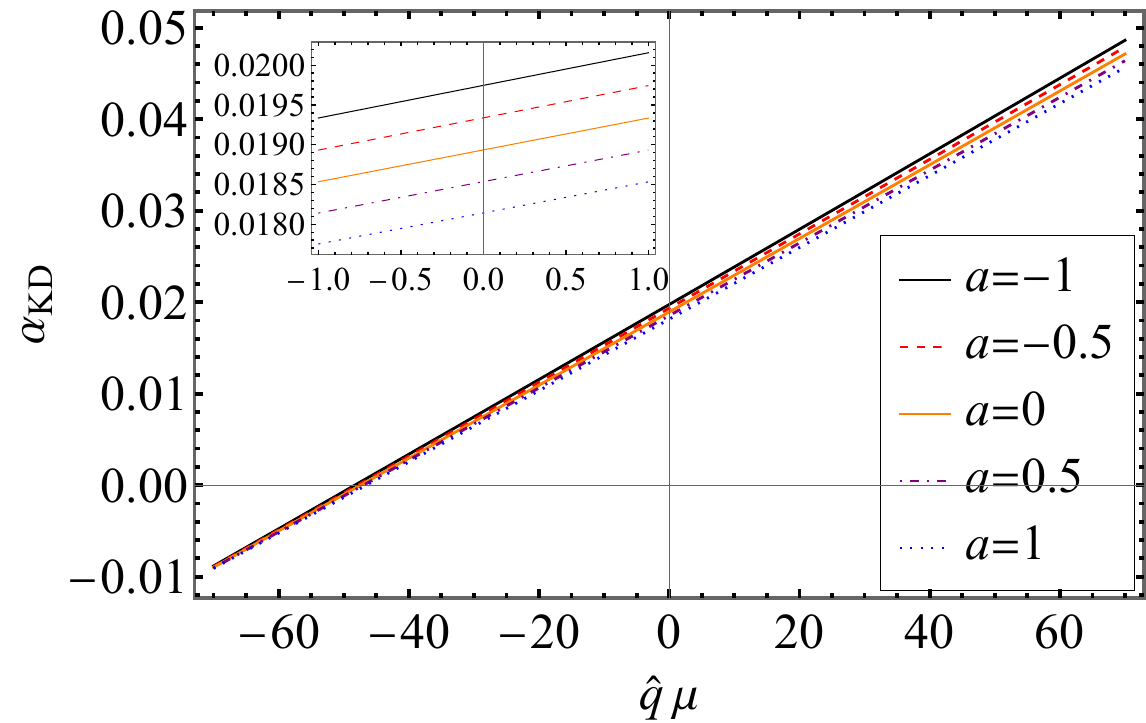}
    \caption{The dependence of PS on $\hat{q}\mu$ of different $a$ with the PN method result Eq. \eqref{eq:PS-Kerr-Dipole} truncated to 6-th order. The other parameters used are $e = 1/10$, $p = 1000M$, $\ M = 1$, $\ s=1$.}
    \label{fig:PN-Parameter}
\end{figure}

\begin{figure}[htp!]
    \centering
    \includegraphics[width=0.45\textwidth]{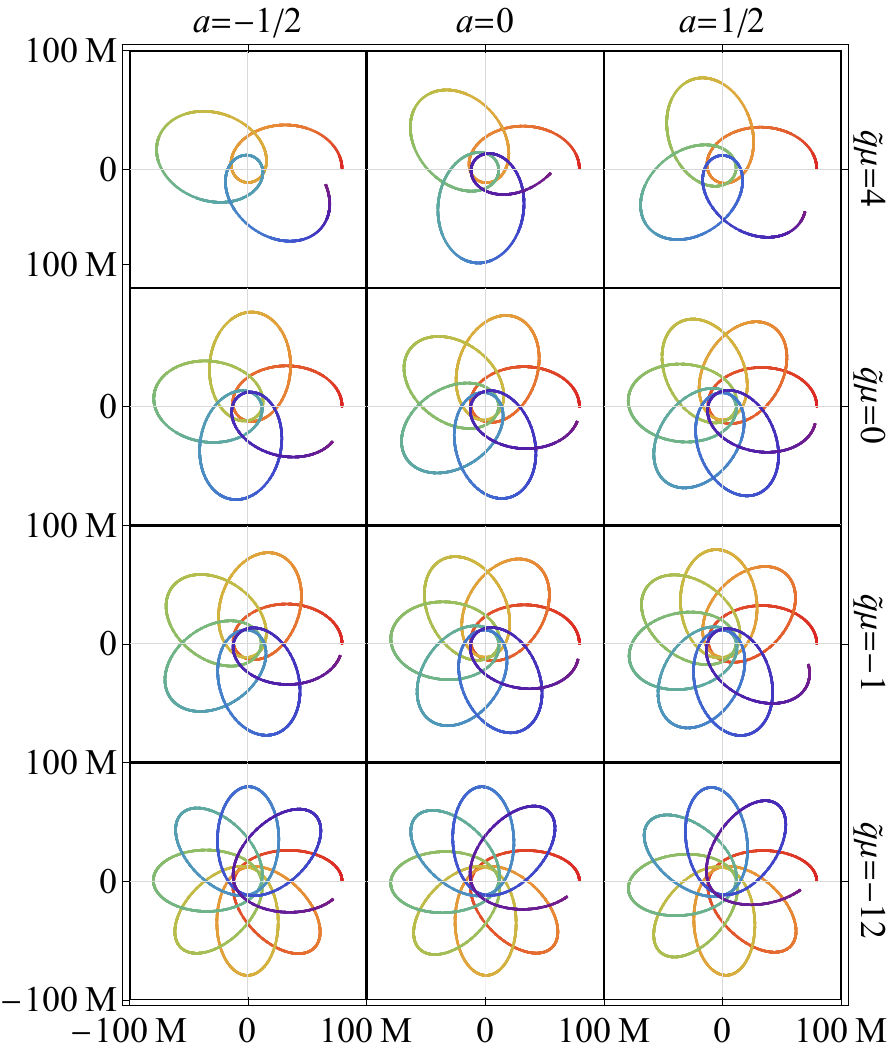}
    \caption{The counterclockwise orbits of charged test particles in dipole magnetized Kerr spacetime with $a$ and $\hat{q}\mu$ being the variables. The starting point is placed on the positive $\mathcal{}{x}$-axis. Other parameters are $p=20M$, $e=3/4$ and $M=1$. }
    \label{fig:orbit in uniformly magnetized Kerr}
\end{figure}

Instead, we study more carefully the magnetic contribution on the PS. 
Note that due to the particular form of the four-potential, i.e., the fact both $\mathcal{A}_t$ and $\mathcal{A}_\phi$ depend linearly on $\mu$, we see from Eq. \eqref{eq:PS-Kerr-Dipole} that $\hatq$ and $\mu$ appear in both the electrostatic and magnetic contributions through and only through their product $\hatq \mu$. Therefore in Fig. \ref{fig:PN-Parameter}, we plot directly the dependence of $\alpha_{\mathrm{KD}}$ on $\hatq\mu$ for a few $a$'s. And in Fig. \ref{fig:orbit in uniformly magnetized Kerr}, we plot the corresponding trajectories for a few choices of $\hatq \mu$ and $a$. To make the PS recognizable in this figure, we chose a smaller $p$ than in Fig. \ref{fig:PN-Parameter}.

It is observed from Fig. \ref{fig:PN-Parameter}.
that when $|\hatq\mu|$ is very small, the PS is positive for the given parameter setting. This agrees with the PS we can recognize from the second row of Fig. \ref{fig:orbit in uniformly magnetized Kerr}. As $\hatq\mu$ grows positive and larger, the magnetic interaction linearly increases the PS, as dictated by the $4\pi\hatq \mu/(\sqrt{M}p^{3/2})$ term in Eq. \eqref{eq:PS-Kerr-Dipole}. This also agrees with the increase of the PS in the first row of  Fig. \ref{fig:orbit in uniformly magnetized Kerr} compared to the second row. When $\hatq\mu$ is negative, however, we see that as $|\hatq\mu|$ increases and eventually larger than $3M^{3/2}\sqrt{p}/2$, the PS could become negative, as shown by the left part of Fig. \ref{fig:PN-Parameter}. Note that $\hatq\mu$ at this value will not spoil the convergence of the series in Eq. \eqref{eq:PS-Kerr-Dipole} because this series can still be convergent every other order under these parameter settings. Physically negative $\alpha_{\mathrm{KD}}$ means that the periapsis shifts backward in the rotation direction of the trajectory, which is solely an effect of the Lorentz magnetic force. When this happens, essentially the magnetic force is so strong that the particle reach the periapsis first before reaching the same azimuth angle, as seen from the last row of Fig. \ref{fig:orbit in uniformly magnetized Kerr}.
From the insert of Fig. \ref{fig:PN-Parameter} we also observe that as $a$ increases, the PS decreases monotonically, as dictated by the FD term. 

\section{Quasi-circular PS in uniform magnetic field}  \label{sec:QPO}

When the magnetic field is asymptotically stronger than the dipole magnetic field, i.e., decays slower than $1/r^3$, the PS can not be computed using the PN method in the previous section because the motion is so dramatically different from a quasi-elliptical motion. For example, for a uniform magnetic field, the orbit at a large radius will be dictated by the Lorentz force to be a quasi-circle. At smaller radii, although the effect of gravitation becomes more important, the PN method also become less accurate because $p$ is not that large. Therefore at both large and small radii, we will need a new method to compute the PS for the magnetic field decaying not fast enough. 
Although a uniform magnetic field seems unlikely in the large scale, it remains one of the simplest case in many theoretical and astronomical models of magnetic field, especially when studying its effect not very far away from the center. Therefore in this section, we will study the PS for a uniform magnetic field, without having to set the radius to a large value.

We will concentrate on the orbits that are at least quasi-circular with radius 
\begin{align}
    r(\tau)=r_0+\delta r(\tau), \label{eq:rinr0}
\end{align} 
where $r_0$ is the circular orbit radius and $\delta r$ is a small perturbation. 
The radius $r_0$ satisfies the conditions 
\begin{align}
    V_{\mathrm{eff}}(r)=0,\, V'_{\mathrm{eff}}(r)=0, \label{eq:r0cond}
\end{align}
where $V_{\mathrm{eff}}$ is given in Eq. \eqref{Veff_SAS}. These conditions effectively establish two relations between the kinetic variables $E,\,L$ and $r_0$. They are two coupled second-order polynomials of $E$ and $L$ whose solutions $E(r_0)$ and $L(r_0)$ can be analytically expressed using elementary functions, although their forms are too lengthy to give here. 
In the following, we will treat $E$ and $L$ are functions of $r_0$ respectively.

Substituting Eq. \eqref{eq:rinr0} into Eq. \eqref{rdot2} and using conditions \eqref{eq:r0cond}, the perturbative function $\delta r$ is found to satisfy the harmonic oscillation equation 
\begin{equation}
    \frac{\dd^2 \delta r}{\dd \tau^2}+\omega_r^2 \delta r=0, 
\end{equation}
where the oscillation frequency is
\begin{equation}
    \omega_r^2=\frac{1}{B}\frac{\partial^2 V_{\mathrm{eff}}}{\partial r^2}\bigg|_{(r=r_0,\theta=\frac{\pi}{2})}.  \label{Radial Frequency Definition}
\end{equation}
Additionally, the orbital angular frequency is given by the following expression.
\begin{equation}
    \omega_{\phi}=\frac{\dd \phi}{\dd\tau}\bigg|_{(r=r_0,\theta=\frac{\pi}{2})}.  \label{Orbital Angular Frequency}
\end{equation}

The PS then is related to $\omega_r$ and $\omega_\phi$ as
\begin{equation}
    \alpha_{\mathrm{QC}}=2\pi\left(\left|\frac{\omega_{\phi}}{\omega_{r}}\right|-1\right),
 \label{eq:HO PS formular}
\end{equation}
where the introduction of the absolute value aims to covering the description of particle motion in both the clockwise and counterclockwise directions.
After substituting Eq. \eqref{Radial Frequency Definition}, Eq. \eqref{Orbital Angular Frequency} and Eq. \eqref{Veff_SAS}, it can be written as 
\begin{align}
\alpha_{\mathrm{QC}}=&2\pi\Bigg\{\Bigg|\frac{A\left(L-\hat{q} \mathcal{A}_{\phi }\right)-D(E+\hat{q} \mathcal{A}_t)}{D^2+AC}\sqrt{2B} \notag \\&\times\bigg(\frac{\partial^2}{\partial r^2}\bigg[\frac{1}{D^2+AC}\Big(C (E+\hat{q} \mathcal{A}_t)^2-A (L-\hat{q} \mathcal{A}_{\phi})^2\notag\\ &+2D (E+\hat{q} \mathcal{A}_t)(L-\hat{q} \mathcal{A}_{\phi})\Big)\bigg]\bigg)^{-\frac{1}{2}}\Bigg|-1\Bigg\}\bigg|_{(r=r_0)}.
\label{eq:aqc}
\end{align}

We now use Eq. \eqref{eq:aqc} to compute the PS in Kerr spacetime with an additional uniform magnetic field along the spin axis of the BH. The nonzero components of the four-potential in this case have the form   \cite{Wald_1974,Aliev_Galtsov_1989,Aliev_Ozdemir_2002,Prasanna_1978,Aliev_Galtsov_1989}
\begin{subequations}
\label{eq:kmagfp}
\begin{align}
    &\mathcal{A}_{t}=a B_0 \left[\frac{M r} {\Sigma} \, ( 1+\cos^2 \theta) -1\right],  \label{Kerr uniform MF A_t explicit}\\
    &\mathcal{A}_{\phi}=B_0 \sin^2\theta\left[\frac{r^2+a^2}{2}-\frac{Mra^2}{\Sigma}( 1+\cos^2 \theta) \right], \label{Kerr uniform MF A_phi explicit}
\end{align}
\end{subequations}
where $B_0$ is the strength of the magnetic field. 

Substituting these equations and the Kerr metric Eq. \eqref{Kerr metric} into Eq. \eqref{eq:aqc}, the PS $\alpha_{\mathrm{KU}}$ in the Kerr spacetime with a uniform magnetic field is obtained as 

\begin{widetext}
\begin{align}
    \alpha_{\mathrm{KU}}=&2 \pi \Bigg\{-1+r_0^{3/2} \Big[(r_0-2M) \left(L-r_0^2 \mathcal{B}\right)+2 a E M-a^2 r_0 \mathcal{B}\Big]\bigg[-3 L^2 r_0^2 (2 M-r_0)^3-2 E^2 M r_0^6\notag\\
    &+ r_0^6 (r_0-2 M)^3\mathcal{B}^2+8 a E L M r_0^2 \left(6 M^2-8 M r_0+3 r_0^2\right)+a^2 r_0 \Big(12 L^2 M^2 +r_0 M\left(-24 E^2 M^2-6 L^2 \right)\notag\\
    &+\left(28 E^2 M^2-L^2\right) r_0^2 -6 E^2 M r_0^3+r_0 (r_0-2 M)^2 \left(-4 M^2+2 M r_0+3 r_0^2\right)\mathcal{B}^2 \Big)+12 a^3 E L M r_0 (r_0-2 M)\notag\\
    &+a^4 \Big(-2 M \left(L^2+3 E^2 r_0 (r_0-2 M)\right)+3 r_0^2 \mathcal{B}^2 \left(8 M^3-8 M^2 r_0+r_0^3\right)\Big) +  4 a^5 E L M+a^6 \Big(-2 E^2 M\notag\\
    &+r_0 \left(-12 M^2+6 M r_0+r_0^2\right)\mathcal{B}^2\Big)+2 a^8 M \mathcal{B}^2\bigg]^{-\frac{1}{2}}\Bigg\},  \label{eq:alpha kerr uniform}
\end{align}    
\end{widetext}
where for convenience we have introduced the notation for the Larmor frequency 
\begin{equation}
    \mathcal{B}=\frac{\hat{q}B_0}{2}.
\end{equation}
Note that this quantity can be quite large even for weak magnetic fields if the specific charge $\hat{q}$ is large, as in the case of electrons.
We emphasis that unlike the PS in the previous section, the PS \eqref{eq:alpha kerr uniform} is valid for $r_0$ that is not necessarily large. 
When the magnetic field is turned off, this PS reduces to that of the quasi-circular orbit in the Kerr spacetime
\begin{widetext}
\begin{align}
    \alpha_{\mathrm{K}}=&2 \pi \Bigg\{-1+r_0^{3/2} \Big[(r_0-2M) L+2 a E M\Big]\bigg[\Big(-3 L^2 r_0^2 (2 M-r_0)^3-2 E^2 M r_0^6 +8 a E L M r_0^2 \left(6 M^2-8 M r_0+3 r_0^2\right)\notag\\
    &+a^2 r_0 \left(L^2 \left(12 M^2-6 M r_0-r_0^2\right)-2 E^2 M r_0 \left(12 M^2-14 M r_0+3 r_0^2\right)\right)+12 a^3 E L M r_0 (r_0-2 M)-2 a^4 M \left(L^2
    \right.\notag\\
    &\left.+3 E^2 r_0 (r_0-2 M)\right)+  4 a^5 E L M-2 a^6 E^2 M \Big)\bigg]^{-\frac{1}{2}}\Bigg\}.  \label{eq:alpha kerr}
\end{align}    
\end{widetext}
After substituting the analytical $E,\ L$ and expanding at large $r_0$, Eq. \eqref{eq:alpha kerr}  coincides with the results Eq. \eqref{eq:PS-Kerr}. 

\begin{figure}[htp!]
    \centering
    \begin{subfigure}{0.5\textwidth}
        \includegraphics[width=0.8\textwidth]{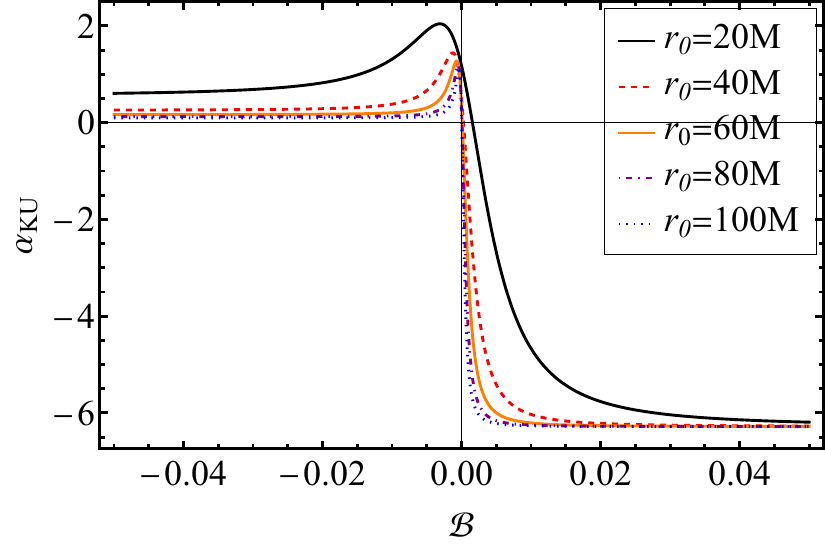}
        \caption{}
        \label{fig:QPO进动角与qB的关系}
    \end{subfigure}
    \begin{subfigure}{0.5\textwidth}
        \includegraphics[width=0.9\textwidth]{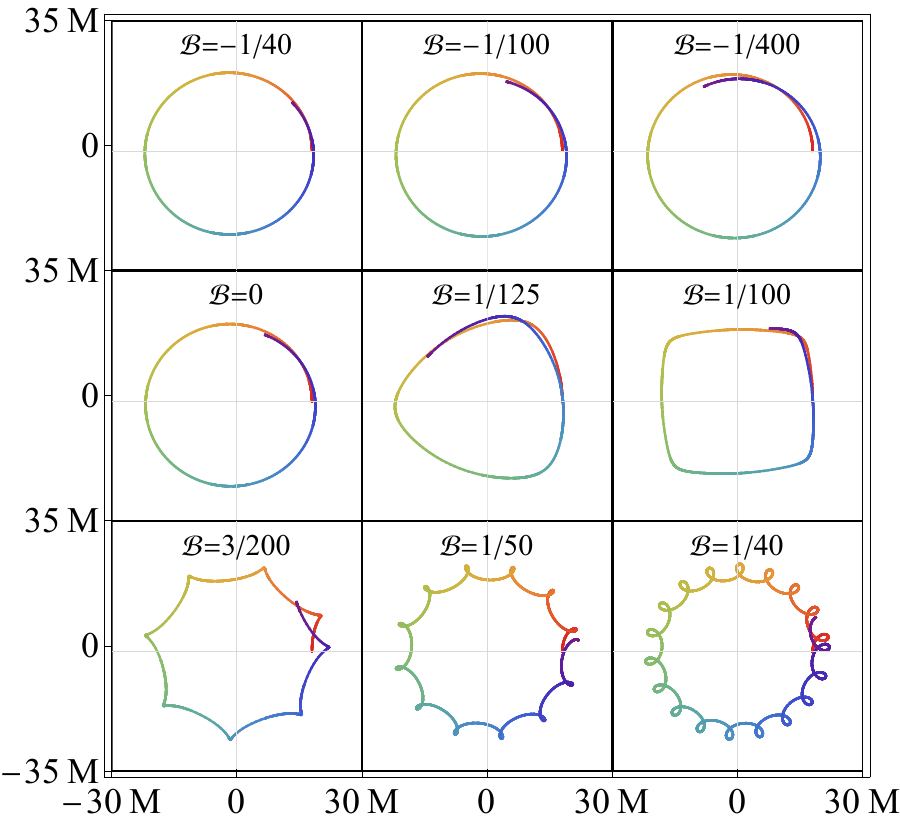}
        \caption{}
        \label{fig:Curledorbit}
    \end{subfigure}
    \caption{Dependence of $\alpha_{\mathrm{KU}}$ Eq. \eqref{eq:alpha kerr uniform} on $a, \mathcal{B}$ and $r$. The default parameters are $\mathcal{B} = 10^{-5}$, $r_0 = 20M $, $\Delta r_0= \pm 2M$, $a = M/10$, $M=1$ except the variables. In Fig. \ref{fig:QPO-PS对参数的依赖性}(c), the first four trajectories start at periapsis and end at second periapsis.}
    \label{fig:QPO-PS对参数的依赖性}
\end{figure}

To study the effect of the magnetic interaction and the spacetime spin on $\alpha_{\mathrm{KU}}$, in Fig. \ref{fig:QPO-PS对参数的依赖性} we plot its dependence on parameters $\mathcal{B}$, as well as the trajectory plot corresponding to some fixed choices of the parameters obtained using a numerical method. 
As illustrated in Fig. \ref{fig:QPO-PS对参数的依赖性} (a), as the repulsive Lorentz force $(\mathcal{B}>0)$ increases, the PS decreases towards the constant value of $-2\pi$. This corresponds to the observation in Fig. \ref{fig:QPO-PS对参数的依赖性} (b) that as $\mathcal{B}$ increases, the trajectories grow cusps and even local helical loops so that the PS defined as the angle between two minimal (or maximal) $r$ minus $2\pi$ eventually is dominated by $-2\pi$. 
In contrast, we see from Fig. \ref{fig:QPO-PS对参数的依赖性} (a) that when the Lorentz force is attractive $(\mathcal{B}<0)$, the PS exhibits a non-monotonic behavior, initially increasing before subsequently decreasing as $|\mathcal{B}|$ increases. This agrees with the three subplots in the first row of Fig. \ref{fig:QPO-PS对参数的依赖性} (b). 

Furthermore, Fig. \ref{fig:QPO-PS对参数的依赖性} (a) demonstrates that, regardless of whether the Lorentz force is repulsive or attractive, the PS generally decreases with increasing $r_0$. 
Finally, 
we also plotted (did but not show) the dependence 
of $\alpha_{\mathrm{KU}}$ on the 
spacetime spin $a$ and find that as it increases, the value of the PS decreases almost linearly. This agrees with the effect of $a$ on the PS of particles in Kerr spacetime without a magnetic field in our previous study \cite{He_2024}, where this decrease with $a$ was attributed to the FD effect.

\section{Application to the PS of Mercury and S2}    \label{sec:application}

In this section, we will apply the PS result Eq. \eqref{eq:PN PS general series expansion} to known systems with measured PS to constrain the magnetic field of the central body and/or charge of the orbiting object. The first such system is the Sun-Mercury system whose PS has been measured for a long time. The second is the recently measured Sgr A* - S$_2$ system whose PS is measured only in recent years. 

\subsection{PS of Mercury and constraints}

A periodic polar magnetic field exists near the solar surface, which reverses approximately every 11 years due to changes in solar activity \cite{Mathew_2013,HaleCycle_2015,Zihao_2024}. 
The magnetic field generated by the Sun is carried by the solar wind to planets' orbits and forms the interplanetary magnetic field. 
During solar minima, the magnetic field near the sun can be approximated as a dipolar magnetic field \cite{Wang_2003,Wang_2009,Mathew_2013,HaleCycle_2015}.
However due to the complexity of solar activity and solar wind, to what extent this dipolar structure is kept at the Mercury orbit is unclear to us. Therefore in this subsection, we will assume that the magnetic field experienced by Mercury is still of a dipolar shape and therefore our results in Subsec. \ref{sec:PN method} can be applied. 

\begin{figure}[htp!]
    \centering
    \begin{subfigure}{0.5\textwidth}
        \includegraphics[width=0.8\textwidth]{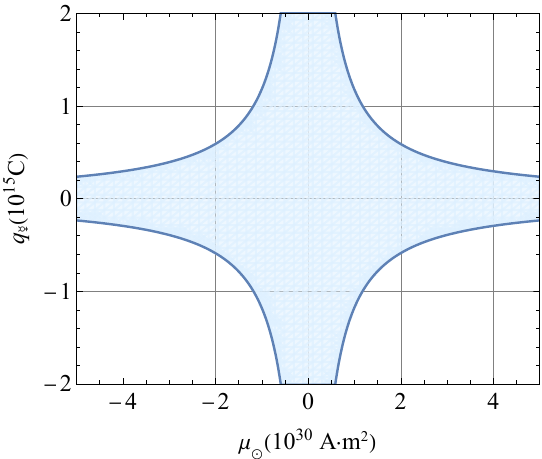}
        \caption{}
        \label{fig:SunConstraint}
    \end{subfigure}
    \begin{subfigure}{0.5\textwidth}
        \includegraphics[width=0.83\textwidth]{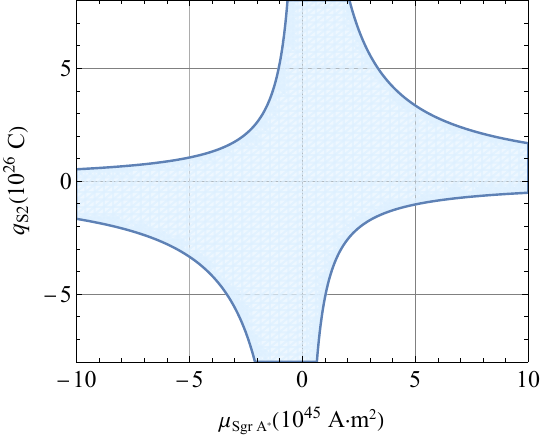}
        \caption{}
        \label{fig:SgrAConstraint}
    \end{subfigure}
    \caption{(a) Allowed parameter space of $q_{\mercury}$ and $\mu _{\odot }$. $M _{\odot}=1.988\times 10^{30}\ \text{kg}$ \cite{Prsa_2016}, $M_{\mercury}=3.301\times 10^{23}\ \text{kg}$ \cite{Luzum_2011}, $e=0.2056$, $p=5.545\times 10^{10}\ \text{m}$ \cite{Park_2017} for Mercury are used. (b) Allowed parameter space of $q_{\text{S2}}$ and $\mu _{\text{Sgr A*}}$. $M _{\text{Sgr A*}}=4.26\times 10^{6} M _{\odot }$ \cite{GRAVITY:2020gka}, $M_{\text{S2}}=13.6 M _{\odot }$ \cite{Habibi_2017}, $e=0.8847$, $p=2.23\times 10^{2}\ \text{AU}$ \cite{GRAVITY:2020gka} are used. }
    \label{fig:Constraint on sun}
\end{figure}

The MESSENGER spacecraft has measured Mercury's precession \cite{Park_2017} with an uncertainty of $1.5\times {10^{-3}}\ ''/ \text{cty} $ after subtracting the Schwarzschild-like and FD contributions. 
If we interpret this uncertainty as the effect of the solar magnetic field on the PS of charged Mercury, we can constrain the dipolar moment $\mu$ and Mercury's charge $Q$. 
The leading order magnetic interaction contribution in Eq. \eqref{eq:PS-Kerr-Dipole} is of the form 
\begin{align}
\Delta\alpha=\frac{4 \pi \hat{q} \mu}{\sqrt{M}p^{3/2}},
\end{align}
and the size of this should be smaller than the uncertainty above, i.e., $|\Delta\alpha|<1.5\times 10^{-3}\ ''/ \text{cty} $. 
Effectively, this inequality constrains the value of $\mu$ and $\hatq_{\mercury}$ to a region in the $(\mu,\,\hatq_{\mercury})$ parameter space bounded by four reciprocal functions, as shown in Fig. \ref{fig:Constraint on sun}. 

Moreover, using the rough magnetic field of 10 $G$ near the Sun's polar region \cite{Wang_2003,Petrie_2022}, we can estimate the dipole moment of the solar magnetic field to be at the scale of $10^{30}\sim 10^{31}~\text{A}\cdot \text{m}^2$. Although these values are not very accurate due to the variation of the solar activity and consequently its magnetic field, we can still use them to order-estimate the value of $q_{\mercury}$ with the help of Fig. \ref{fig:Constraint on sun}. That is, we can constrain the value of $|q_{\mercury}|<10^{15}\sim 10^{16}~\text{C}$.

\subsection{PS of S2 and constraints}

In 2020, the GRAVITY Collaboration reported the detection of the Schwarzschild precession of star S2 orbiting the massive compact radio source Sgr A*, yielding a dimensionless parameter of $f_{\text{SP}} = 1.10 \pm 0.19$ to describe the deviation of detected PS value from the  Schwarzschild precession \cite{GRAVITY:2020gka}. 

If we attribute this deviation to the effect of the magnetic field associated with Sgr A* on the charge of S2, we can constrain the magnetic field of Sgr A* and the charge of S2 in a similar way as for Mercury. 
The magnetic field associated with Sgr A* was studied in the accretion processes around Sgr A* in Ref. \cite{eatough_strong_2013}, based on which we can also assume that the magnetic field around Sgr A* is of a dipolar form. 

Assuming that the deviation of $f_{\mathrm{SP}}-1\in[-0.09,\,0.29]$ is completely due to the dipole magnetic field interacting with S2 charge, then using the leading term of this contribution, i.e., also the term 
\begin{align}
\Delta\alpha=\frac{4 \pi \hatq_{\mathrm{S2}} \mu_{\mathrm{Sgr\ A^*}}}{\sqrt{M_{\mathrm{Sgr\ A^*}}}p_{\mathrm{S2}}^{3/2}},
\end{align}
we can find the allowed region in the $(\mu_{\mathrm{Sgr\ A^*}},\, q_{\mathrm{S2}})$ parameter space, as illustrated in Fig. \ref{fig:Constraint on sun} (b). 

\section{conclusion}    \label{sec:conclusion}

In this study, we investigated the PS of charged test particles in the equatorial plane of general SAS spacetimes with axisymmetric magnetic fields. For magnetic fields that decay fast enough, such as those with non-zero dipole or higher order moment, it is found that the PS can be computed using the PN method to high orders of $p$ for arbitrary $e$, with the final results given in Eq. \eqref{eq:PS-Kerr-Dipole}. For magnetic fields that decay more slowly however, such as the asymptotically uniform magnetic field, we found that the PS in general can not be computed using the PN method with general $e$ since the trajectories do not assemble the elliptical form. Instead, we found the PS for quasi-circular orbits with arbitrary radius $r_0$. The general result for this case is given in Eq. \eqref{eq:alpha kerr uniform}.

The former case above was applied to the Kerr spacetime with a dipolar magnetic field and the PS result is given in Eq. \eqref{eq:PS-Kerr-Dipole} and its various limits are obtained. Moreover, we verified the PN PS in this case with numerical integration results and excellent agreement was found. The PS in this case is found to receive the magnetic correction proportional to $q\mu/m$ starting from the $p^{3/2}$ order, the same order as the FD effect to the PS in \cite{He_2024}. At this order, the magnetic field increases (or decreases) the PS if the Lorentz force is attractive (or repulsive). Moreover, this contribution can in principle change the sign of the total PS from positive, i..e, the periapsis point is reached again after rotating one full turn, to negative when the periapsis is reached before reaching the same azimuth angle. 
For the PS of quasi-circular orbit with uniform field, the general result is presented in Eq. \eqref{eq:alpha kerr uniform}. Note that this result did not use the large radius approximation and therefore allows us to study the PS when the orbit radius is small. The effect of the magnetic field in this case is non-monotonic, as shown in Fig. \ref{fig:QPO-PS对参数的依赖性}: when the Lorentz force is repulsive and increasing, the quasi-circular PS decreases from a positive value to $-2\pi$. While for an attractive Lorentz force, the quasi-circular PS increases first and then approaches zero as the attractive force increases. 

The PS for a dipolar magnetic field is used to constrain the charge of the test particle and the dipole moment of the magnetic field in two systems, the Sun-Mercury and the Sgr A*-S2 systems, as seen in Fig. \ref{fig:Constraint on sun}. 
This suggests that by analyzing the modifications to the precession angle induced by the Lorentz force acting on charged particles, we can extract significant information regarding the black hole's magnetic field.

\section*{Acknowledgments}

We thank the discussion with Jinhong He and Qianchuan Wang. This work is partially supported by a research development fund from Wuhan University. 


\begin{thebibliography}{56}%
\makeatletter
\providecommand \@ifxundefined [1]{%
 \@ifx{#1\undefined}
}%
\providecommand \@ifnum [1]{%
 \ifnum #1\expandafter \@firstoftwo
 \else \expandafter \@secondoftwo
 \fi
}%
\providecommand \@ifx [1]{%
 \ifx #1\expandafter \@firstoftwo
 \else \expandafter \@secondoftwo
 \fi
}%
\providecommand \natexlab [1]{#1}%
\providecommand \enquote  [1]{``#1''}%
\providecommand \bibnamefont  [1]{#1}%
\providecommand \bibfnamefont [1]{#1}%
\providecommand \citenamefont [1]{#1}%
\providecommand \href@noop [0]{\@secondoftwo}%
\providecommand \href [0]{\begingroup \@sanitize@url \@href}%
\providecommand \@href[1]{\@@startlink{#1}\@@href}%
\providecommand \@@href[1]{\endgroup#1\@@endlink}%
\providecommand \@sanitize@url [0]{\catcode `\\12\catcode `\$12\catcode `\&12\catcode `\#12\catcode `\^12\catcode `\_12\catcode `\%12\relax}%
\providecommand \@@startlink[1]{}%
\providecommand \@@endlink[0]{}%
\providecommand \url  [0]{\begingroup\@sanitize@url \@url }%
\providecommand \@url [1]{\endgroup\@href {#1}{\urlprefix }}%
\providecommand \urlprefix  [0]{URL }%
\providecommand \Eprint [0]{\href }%
\providecommand \doibase [0]{https://doi.org/}%
\providecommand \selectlanguage [0]{\@gobble}%
\providecommand \bibinfo  [0]{\@secondoftwo}%
\providecommand \bibfield  [0]{\@secondoftwo}%
\providecommand \translation [1]{[#1]}%
\providecommand \BibitemOpen [0]{}%
\providecommand \bibitemStop [0]{}%
\providecommand \bibitemNoStop [0]{.\EOS\space}%
\providecommand \EOS [0]{\spacefactor3000\relax}%
\providecommand \BibitemShut  [1]{\csname bibitem#1\endcsname}%
\let\auto@bib@innerbib\@empty
\bibitem [{\citenamefont {Einstein}(1916)}]{Einstein:1916}%
  \BibitemOpen
  \bibfield  {author} {\bibinfo {author} {\bibfnamefont {A.}~\bibnamefont {Einstein}},\ }\href {https://doi.org/https://doi.org/10.1002/andp.19163540702} {\bibfield  {journal} {\bibinfo  {journal} {Annalen der Physik}\ }\textbf {\bibinfo {volume} {354}},\ \bibinfo {pages} {769} (\bibinfo {year} {1916})}\BibitemShut {NoStop}%
\bibitem [{\citenamefont {Lucchesi}\ and\ \citenamefont {Peron}(2010)}]{Lucchesi_2010}%
  \BibitemOpen
  \bibfield  {author} {\bibinfo {author} {\bibfnamefont {D.~M.}\ \bibnamefont {Lucchesi}}\ and\ \bibinfo {author} {\bibfnamefont {R.}~\bibnamefont {Peron}},\ }\href {https://doi.org/10.1103/PhysRevLett.105.231103} {\bibfield  {journal} {\bibinfo  {journal} {Phys. Rev. Lett.}\ }\textbf {\bibinfo {volume} {105}},\ \bibinfo {pages} {231103} (\bibinfo {year} {2010})}\BibitemShut {NoStop}%
\bibitem [{\citenamefont {Everitt}\ \emph {\textit{et~al.}}(2011)\citenamefont {Everitt}, \citenamefont {DeBra}, \citenamefont {Parkinson}, \citenamefont {Turneaure}, \citenamefont {Conklin}, \citenamefont {Heifetz}, \citenamefont {Keiser}, \citenamefont {Silbergleit}, \citenamefont {Holmes}, \citenamefont {Kolodziejczak} \emph {\textit{et~al.}}}]{Everitt_2011}%
  \BibitemOpen
  \bibfield  {author} {\bibinfo {author} {\bibfnamefont {C.~W.~F.}\ \bibnamefont {Everitt}}, \bibinfo {author} {\bibfnamefont {D.~B.}\ \bibnamefont {DeBra}}, \bibinfo {author} {\bibfnamefont {B.~W.}\ \bibnamefont {Parkinson}}, \bibinfo {author} {\bibfnamefont {J.~P.}\ \bibnamefont {Turneaure}}, \bibinfo {author} {\bibfnamefont {J.~W.}\ \bibnamefont {Conklin}}, \bibinfo {author} {\bibfnamefont {M.~I.}\ \bibnamefont {Heifetz}}, \bibinfo {author} {\bibfnamefont {G.~M.}\ \bibnamefont {Keiser}}, \bibinfo {author} {\bibfnamefont {A.~S.}\ \bibnamefont {Silbergleit}}, \bibinfo {author} {\bibfnamefont {T.}~\bibnamefont {Holmes}}, \bibinfo {author} {\bibfnamefont {J.}~\bibnamefont {Kolodziejczak}}, \textit{et~al.},\ }\href {https://doi.org/10.1103/PhysRevLett.106.221101} {\bibfield  {journal} {\bibinfo  {journal} {Phys. Rev. Lett.}\ }\textbf {\bibinfo {volume} {106}},\ \bibinfo {pages} {221101} (\bibinfo {year} {2011})}\BibitemShut {NoStop}%
\bibitem [{\citenamefont {Iorio}(2019)}]{Iorio_2019}%
  \BibitemOpen
  \bibfield  {author} {\bibinfo {author} {\bibfnamefont {L.}~\bibnamefont {Iorio}},\ }\href {https://doi.org/10.3847/1538-3881/ab19bf} {\bibfield  {journal} {\bibinfo  {journal} {The Astronomical Journal}\ }\textbf {\bibinfo {volume} {157}},\ \bibinfo {pages} {220} (\bibinfo {year} {2019})}\BibitemShut {NoStop}%
\bibitem [{\citenamefont {Damour}\ and\ \citenamefont {Schäeer}(1988)}]{Damour_1988}%
  \BibitemOpen
  \bibfield  {author} {\bibinfo {author} {\bibfnamefont {T.}~\bibnamefont {Damour}}\ and\ \bibinfo {author} {\bibfnamefont {G.}~\bibnamefont {Schäeer}},\ }\href {https://doi.org/10.1007/BF02828697} {\bibfield  {journal} {\bibinfo  {journal} {Il Nuovo Cimento B}\ }\textbf {\bibinfo {volume} {101}},\ \bibinfo {pages} {127} (\bibinfo {year} {1988})}\BibitemShut {NoStop}%
\bibitem [{\citenamefont {Abuter}\ \emph {\textit{et~al.}}(2020)\citenamefont {Abuter} \emph {\textit{et~al.}}}]{GRAVITY:2020gka}%
  \BibitemOpen
  \bibfield  {author} {\bibinfo {author} {\bibfnamefont {R.}~\bibnamefont {Abuter}} \textit{et~al.} (\bibinfo {collaboration} {GRAVITY}),\ }\href {https://doi.org/10.1051/0004-6361/202037813} {\bibfield  {journal} {\bibinfo  {journal} {Astron. Astrophys.}\ }\textbf {\bibinfo {volume} {636}},\ \bibinfo {pages} {L5} (\bibinfo {year} {2020})}\BibitemShut {NoStop}%
\bibitem [{\citenamefont {Bini}\ \emph {\textit{et~al.}}(2005)\citenamefont {Bini}, \citenamefont {De~Paolis}, \citenamefont {Geralico}, \citenamefont {Ingrosso},\ and\ \citenamefont {Nucita}}]{Bini_2005}%
  \BibitemOpen
  \bibfield  {author} {\bibinfo {author} {\bibfnamefont {D.}~\bibnamefont {Bini}}, \bibinfo {author} {\bibfnamefont {F.}~\bibnamefont {De~Paolis}}, \bibinfo {author} {\bibfnamefont {A.}~\bibnamefont {Geralico}}, \bibinfo {author} {\bibfnamefont {G.}~\bibnamefont {Ingrosso}},\ and\ \bibinfo {author} {\bibfnamefont {A.}~\bibnamefont {Nucita}},\ }\href {https://doi.org/10.1007/s10714-005-0109-9} {\bibfield  {journal} {\bibinfo  {journal} {General Relativity and Gravitation}\ }\textbf {\bibinfo {volume} {37}},\ \bibinfo {pages} {1263–1276} (\bibinfo {year} {2005})}\BibitemShut {NoStop}%
\bibitem [{\citenamefont {Vogt}\ and\ \citenamefont {Letelier}(2008)}]{Vogt_2008}%
  \BibitemOpen
  \bibfield  {author} {\bibinfo {author} {\bibfnamefont {D.}~\bibnamefont {Vogt}}\ and\ \bibinfo {author} {\bibfnamefont {P.~S.}\ \bibnamefont {Letelier}},\ }\href {https://doi.org/10.1111/j.1365-2966.2007.12771.x} {\bibfield  {journal} {\bibinfo  {journal} {Monthly Notices of the Royal Astronomical Society}\ }\textbf {\bibinfo {volume} {384}},\ \bibinfo {pages} {834} (\bibinfo {year} {2008})}\BibitemShut {NoStop}%
\bibitem [{\citenamefont {Schmidt}(2011)}]{Schmidt_2011}%
  \BibitemOpen
  \bibfield  {author} {\bibinfo {author} {\bibfnamefont {H.-J.}\ \bibnamefont {Schmidt}},\ }\href {https://doi.org/10.1103/PhysRevD.83.124010} {\bibfield  {journal} {\bibinfo  {journal} {Phys. Rev. D}\ }\textbf {\bibinfo {volume} {83}},\ \bibinfo {pages} {124010} (\bibinfo {year} {2011})}\BibitemShut {NoStop}%
\bibitem [{\citenamefont {Shchigolev}(2016)}]{Shchigolev_2016}%
  \BibitemOpen
  \bibfield  {author} {\bibinfo {author} {\bibfnamefont {V.}~\bibnamefont {Shchigolev}},\ }\href {https://doi.org/10.14419/ijpr.v4i2.6530} {\bibfield  {journal} {\bibinfo  {journal} {International Journal of Physical Research}\ }\textbf {\bibinfo {volume} {4}},\ \bibinfo {pages} {52} (\bibinfo {year} {2016})}\BibitemShut {NoStop}%
\bibitem [{\citenamefont {Mak}\ \emph {\textit{et~al.}}(2018)\citenamefont {Mak}, \citenamefont {Leung},\ and\ \citenamefont {Harko}}]{Mak_2018}%
  \BibitemOpen
  \bibfield  {author} {\bibinfo {author} {\bibfnamefont {M.~K.}\ \bibnamefont {Mak}}, \bibinfo {author} {\bibfnamefont {C.~S.}\ \bibnamefont {Leung}},\ and\ \bibinfo {author} {\bibfnamefont {T.}~\bibnamefont {Harko}},\ }\href {https://doi.org/10.1155/2018/7093592} {\bibfield  {journal} {\bibinfo  {journal} {Advances in High Energy Physics}\ }\textbf {\bibinfo {volume} {2018}},\ \bibinfo {pages} {1–15} (\bibinfo {year} {2018})}\BibitemShut {NoStop}%
\bibitem [{\citenamefont {Walters}(2018)}]{Walters_2018}%
  \BibitemOpen
  \bibfield  {author} {\bibinfo {author} {\bibfnamefont {S.~J.}\ \bibnamefont {Walters}},\ }\href {https://doi.org/10.1093/mnras/sty2101} {\bibfield  {journal} {\bibinfo  {journal} {Monthly Notices of the Royal Astronomical Society}\ }\textbf {\bibinfo {volume} {480}},\ \bibinfo {pages} {3747} (\bibinfo {year} {2018})}\BibitemShut {NoStop}%
\bibitem [{\citenamefont {Cao}\ \emph {\textit{et~al.}}(2017)\citenamefont {Cao}, \citenamefont {Gong}, \citenamefont {Meng}, \citenamefont {Xu}, \citenamefont {Chen}, \citenamefont {Guo}, \citenamefont {Li}, \citenamefont {Liu}, \citenamefont {Xue}, \citenamefont {Cao} \emph {\textit{et~al.}}}]{Cao_2017}%
  \BibitemOpen
  \bibfield  {author} {\bibinfo {author} {\bibfnamefont {Y.}~\bibnamefont {Cao}}, \bibinfo {author} {\bibfnamefont {Y.}~\bibnamefont {Gong}}, \bibinfo {author} {\bibfnamefont {X.}~\bibnamefont {Meng}}, \bibinfo {author} {\bibfnamefont {C.~K.}\ \bibnamefont {Xu}}, \bibinfo {author} {\bibfnamefont {X.}~\bibnamefont {Chen}}, \bibinfo {author} {\bibfnamefont {Q.}~\bibnamefont {Guo}}, \bibinfo {author} {\bibfnamefont {R.}~\bibnamefont {Li}}, \bibinfo {author} {\bibfnamefont {D.}~\bibnamefont {Liu}}, \bibinfo {author} {\bibfnamefont {Y.}~\bibnamefont {Xue}}, \bibinfo {author} {\bibfnamefont {L.}~\bibnamefont {Cao}}, \textit{et~al.},\ }\href {https://api.semanticscholar.org/CorpusID:119237090} {\bibfield  {journal} {\bibinfo  {journal} {Monthly Notices of the Royal Astronomical Society}\ } (\bibinfo {year} {2017})}\BibitemShut {NoStop}%
\bibitem [{\citenamefont {Shchigolev}(2015)}]{Shchigolev_2015}%
  \BibitemOpen
  \bibfield  {author} {\bibinfo {author} {\bibfnamefont {V.~K.}\ \bibnamefont {Shchigolev}},\ }\href {https://doi.org/10.13189/ujcmj.2015.030401} {\bibfield  {journal} {\bibinfo  {journal} {Universal Journal of Computational Mathematics}\ }\textbf {\bibinfo {volume} {3}},\ \bibinfo {pages} {45–49} (\bibinfo {year} {2015})}\BibitemShut {NoStop}%
\bibitem [{\citenamefont {Avalos-Vargas}\ and\ \citenamefont {Ares~de Parga}(2012)}]{Avalos_2012}%
  \BibitemOpen
  \bibfield  {author} {\bibinfo {author} {\bibfnamefont {A.}~\bibnamefont {Avalos-Vargas}}\ and\ \bibinfo {author} {\bibfnamefont {G.}~\bibnamefont {Ares~de Parga}},\ }\href {https://doi.org/10.1140/epjp/i2012-12155-2} {\bibfield  {journal} {\bibinfo  {journal} {The European Physical Journal Plus}\ }\textbf {\bibinfo {volume} {127}},\ \bibinfo {pages} {155} (\bibinfo {year} {2012})}\BibitemShut {NoStop}%
\bibitem [{\citenamefont {Hu}\ \emph {\textit{et~al.}}(2013)\citenamefont {Hu}, \citenamefont {Zhang}, \citenamefont {Hou},\ and\ \citenamefont {Tang}}]{Yapeng_2013}%
  \BibitemOpen
  \bibfield  {author} {\bibinfo {author} {\bibfnamefont {Y.}~\bibnamefont {Hu}}, \bibinfo {author} {\bibfnamefont {H.}~\bibnamefont {Zhang}}, \bibinfo {author} {\bibfnamefont {J.-P.}\ \bibnamefont {Hou}},\ and\ \bibinfo {author} {\bibfnamefont {L.-Z.}\ \bibnamefont {Tang}},\ }\href {https://api.semanticscholar.org/CorpusID:55484279} {\bibfield  {journal} {\bibinfo  {journal} {Advances in High Energy Physics}\ }\textbf {\bibinfo {volume} {2014}},\ \bibinfo {pages} {604321} (\bibinfo {year} {2013})}\BibitemShut {NoStop}%
\bibitem [{\citenamefont {Hall}(2022)}]{Michael_2022}%
  \BibitemOpen
  \bibfield  {author} {\bibinfo {author} {\bibfnamefont {M.~J.~W.}\ \bibnamefont {Hall}},\ }\href {https://doi.org/10.1119/5.0098846} {\bibfield  {journal} {\bibinfo  {journal} {American Journal of Physics}\ }\textbf {\bibinfo {volume} {90}},\ \bibinfo {pages} {857} (\bibinfo {year} {2022})}\BibitemShut {NoStop}%
\bibitem [{\citenamefont {Brill}\ and\ \citenamefont {Goel}(1999)}]{Brill_1999}%
  \BibitemOpen
  \bibfield  {author} {\bibinfo {author} {\bibfnamefont {D.~R.}\ \bibnamefont {Brill}}\ and\ \bibinfo {author} {\bibfnamefont {D.}~\bibnamefont {Goel}},\ }\href {https://doi.org/10.1119/1.19255} {\bibfield  {journal} {\bibinfo  {journal} {American Journal of Physics}\ }\textbf {\bibinfo {volume} {67}},\ \bibinfo {pages} {316} (\bibinfo {year} {1999})}\BibitemShut {NoStop}%
\bibitem [{\citenamefont {Colistete~Jr}\ \emph {\textit{et~al.}}(2002)\citenamefont {Colistete~Jr}, \citenamefont {Leygnac},\ and\ \citenamefont {Kerner}}]{Colistete_2002}%
  \BibitemOpen
  \bibfield  {author} {\bibinfo {author} {\bibfnamefont {R.}~\bibnamefont {Colistete~Jr}}, \bibinfo {author} {\bibfnamefont {C.}~\bibnamefont {Leygnac}},\ and\ \bibinfo {author} {\bibfnamefont {R.}~\bibnamefont {Kerner}},\ }\href {https://doi.org/10.1088/0264-9381/19/17/309} {\bibfield  {journal} {\bibinfo  {journal} {Classical and Quantum Gravity}\ }\textbf {\bibinfo {volume} {19}},\ \bibinfo {pages} {4573–4589} (\bibinfo {year} {2002})}\BibitemShut {NoStop}%
\bibitem [{\citenamefont {Jiang}\ and\ \citenamefont {Lin}(2014)}]{jiang_post-newtonian_2014}%
  \BibitemOpen
  \bibfield  {author} {\bibinfo {author} {\bibfnamefont {C.}~\bibnamefont {Jiang}}\ and\ \bibinfo {author} {\bibfnamefont {W.}~\bibnamefont {Lin}},\ }\href {https://doi.org/10.1140/epjp/i2014-14200-6} {\bibfield  {journal} {\bibinfo  {journal} {The European Physical Journal Plus}\ }\textbf {\bibinfo {volume} {129}},\ \bibinfo {pages} {200} (\bibinfo {year} {2014})}\BibitemShut {NoStop}%
\bibitem [{\citenamefont {Heydari-Fard}\ \emph {\textit{et~al.}}(2019)\citenamefont {Heydari-Fard}, \citenamefont {Heydari-Fard},\ and\ \citenamefont {Sepangi}}]{Heydari_Fard_2019}%
  \BibitemOpen
  \bibfield  {author} {\bibinfo {author} {\bibfnamefont {M.}~\bibnamefont {Heydari-Fard}}, \bibinfo {author} {\bibfnamefont {M.}~\bibnamefont {Heydari-Fard}},\ and\ \bibinfo {author} {\bibfnamefont {H.~R.}\ \bibnamefont {Sepangi}},\ }\href {https://doi.org/10.1007/s10714-019-2557-7} {\bibfield  {journal} {\bibinfo  {journal} {General Relativity and Gravitation}\ }\textbf {\bibinfo {volume} {51}},\ \bibinfo {pages} {77} (\bibinfo {year} {2019})}\BibitemShut {NoStop}%
\bibitem [{\citenamefont {Dovciak}\ \emph {\textit{et~al.}}(2004)\citenamefont {Dovciak}, \citenamefont {Karas},\ and\ \citenamefont {Matt}}]{Dovciak_2004}%
  \BibitemOpen
  \bibfield  {author} {\bibinfo {author} {\bibfnamefont {M.}~\bibnamefont {Dovciak}}, \bibinfo {author} {\bibfnamefont {V.}~\bibnamefont {Karas}},\ and\ \bibinfo {author} {\bibfnamefont {G.}~\bibnamefont {Matt}},\ }\href {https://doi.org/10.1111/j.1365-2966.2004.08396.x} {\bibfield  {journal} {\bibinfo  {journal} {Monthly Notices of the Royal Astronomical Society}\ }\textbf {\bibinfo {volume} {355}},\ \bibinfo {pages} {1005–1009} (\bibinfo {year} {2004})}\BibitemShut {NoStop}%
\bibitem [{\citenamefont {Zajaček}\ \emph {\textit{et~al.}}(2019)\citenamefont {Zajaček}, \citenamefont {Tursunov}, \citenamefont {Eckart} \emph {\textit{et~al.}}}]{Zaja_ek_2019}%
  \BibitemOpen
  \bibfield  {author} {\bibinfo {author} {\bibfnamefont {M.}~\bibnamefont {Zajaček}}, \bibinfo {author} {\bibfnamefont {A.}~\bibnamefont {Tursunov}}, \bibinfo {author} {\bibfnamefont {A.}~\bibnamefont {Eckart}}, \textit{et~al.},\ }\href {https://doi.org/10.1088/1742-6596/1258/1/012031} {\bibfield  {journal} {\bibinfo  {journal} {Journal of Physics: Conference Series}\ }\textbf {\bibinfo {volume} {1258}},\ \bibinfo {pages} {012031} (\bibinfo {year} {2019})}\BibitemShut {NoStop}%
\bibitem [{\citenamefont {Daly}(2019)}]{Daly_2019}%
  \BibitemOpen
  \bibfield  {author} {\bibinfo {author} {\bibfnamefont {R.~A.}\ \bibnamefont {Daly}},\ }\href {https://doi.org/10.3847/1538-4357/ab35e6} {\bibfield  {journal} {\bibinfo  {journal} {The Astrophysical Journal}\ }\textbf {\bibinfo {volume} {886}},\ \bibinfo {pages} {37} (\bibinfo {year} {2019})}\BibitemShut {NoStop}%
\bibitem [{\citenamefont {You}\ \emph {\textit{et~al.}}(2023)\citenamefont {You}, \citenamefont {Cao}, \citenamefont {Yan}, \citenamefont {Hameury}, \citenamefont {Czerny}, \citenamefont {Wu}, \citenamefont {Xia}, \citenamefont {Sikora}, \citenamefont {Zhang}, \citenamefont {Du} \emph {\textit{et~al.}}}]{YouBei_2023}%
  \BibitemOpen
  \bibfield  {author} {\bibinfo {author} {\bibfnamefont {B.}~\bibnamefont {You}}, \bibinfo {author} {\bibfnamefont {X.}~\bibnamefont {Cao}}, \bibinfo {author} {\bibfnamefont {Z.}~\bibnamefont {Yan}}, \bibinfo {author} {\bibfnamefont {J.-M.}\ \bibnamefont {Hameury}}, \bibinfo {author} {\bibfnamefont {B.}~\bibnamefont {Czerny}}, \bibinfo {author} {\bibfnamefont {Y.}~\bibnamefont {Wu}}, \bibinfo {author} {\bibfnamefont {T.}~\bibnamefont {Xia}}, \bibinfo {author} {\bibfnamefont {M.}~\bibnamefont {Sikora}}, \bibinfo {author} {\bibfnamefont {S.-N.}\ \bibnamefont {Zhang}}, \bibinfo {author} {\bibfnamefont {P.}~\bibnamefont {Du}}, \textit{et~al.},\ }\href {https://doi.org/10.1126/science.abo4504} {\bibfield  {journal} {\bibinfo  {journal} {Science}\ }\textbf {\bibinfo {volume} {381}},\ \bibinfo {pages} {961} (\bibinfo {year} {2023})}\BibitemShut {NoStop}%
\bibitem [{\citenamefont {Eatough}\ \emph {\textit{et~al.}}(2013)\citenamefont {Eatough}, \citenamefont {Falcke}, \citenamefont {Karuppusamy}, \citenamefont {Lee}, \citenamefont {Champion}, \citenamefont {Keane}, \citenamefont {Desvignes}, \citenamefont {Schnitzeler}, \citenamefont {Spitler}, \citenamefont {Kramer} \emph {\textit{et~al.}}}]{eatough_strong_2013}%
  \BibitemOpen
  \bibfield  {author} {\bibinfo {author} {\bibfnamefont {R.~P.}\ \bibnamefont {Eatough}}, \bibinfo {author} {\bibfnamefont {H.}~\bibnamefont {Falcke}}, \bibinfo {author} {\bibfnamefont {R.}~\bibnamefont {Karuppusamy}}, \bibinfo {author} {\bibfnamefont {K.~J.}\ \bibnamefont {Lee}}, \bibinfo {author} {\bibfnamefont {D.~J.}\ \bibnamefont {Champion}}, \bibinfo {author} {\bibfnamefont {E.~F.}\ \bibnamefont {Keane}}, \bibinfo {author} {\bibfnamefont {G.}~\bibnamefont {Desvignes}}, \bibinfo {author} {\bibfnamefont {D.~H. F.~M.}\ \bibnamefont {Schnitzeler}}, \bibinfo {author} {\bibfnamefont {L.~G.}\ \bibnamefont {Spitler}}, \bibinfo {author} {\bibfnamefont {M.}~\bibnamefont {Kramer}}, \textit{et~al.},\ }\href {https://doi.org/10.1038/nature12499} {\bibfield  {journal} {\bibinfo  {journal} {Nature}\ }\textbf {\bibinfo {volume} {501}},\ \bibinfo {pages} {391} (\bibinfo {year} {2013})}\BibitemShut {NoStop}%
\bibitem [{\citenamefont {Himwich}\ \emph {\textit{et~al.}}(2020)\citenamefont {Himwich}, \citenamefont {Johnson}, \citenamefont {Lupsasca},\ and\ \citenamefont {Strominger}}]{Himwich_2020}%
  \BibitemOpen
  \bibfield  {author} {\bibinfo {author} {\bibfnamefont {E.}~\bibnamefont {Himwich}}, \bibinfo {author} {\bibfnamefont {M.~D.}\ \bibnamefont {Johnson}}, \bibinfo {author} {\bibfnamefont {A.}~\bibnamefont {Lupsasca}},\ and\ \bibinfo {author} {\bibfnamefont {A.}~\bibnamefont {Strominger}},\ }\href {https://doi.org/10.1103/PhysRevD.101.084020} {\bibfield  {journal} {\bibinfo  {journal} {Phys. Rev. D}\ }\textbf {\bibinfo {volume} {101}},\ \bibinfo {pages} {084020} (\bibinfo {year} {2020})}\BibitemShut {NoStop}%
\bibitem [{\citenamefont {{EHT Collaboration}}\ \emph {\textit{et~al.}}(2021{\natexlab{a}})\citenamefont {{EHT Collaboration}} \emph {\textit{et~al.}}}]{EHT_M87_Telescope2}%
  \BibitemOpen
  \bibfield  {author} {\bibinfo {author} {\bibnamefont {{EHT Collaboration}}} \textit{et~al.},\ }\href {https://doi.org/10.3847/2041-8213/abe71d} {\bibfield  {journal} {\bibinfo  {journal} {The Astrophysical Journal Letters}\ }\textbf {\bibinfo {volume} {910}},\ \bibinfo {pages} {L12} (\bibinfo {year} {2021}{\natexlab{a}})}\BibitemShut {NoStop}%
\bibitem [{\citenamefont {{EHT Collaboration}}\ \emph {\textit{et~al.}}(2021{\natexlab{b}})\citenamefont {{EHT Collaboration}} \emph {\textit{et~al.}}}]{EHT_M87_Telescope3}%
  \BibitemOpen
  \bibfield  {author} {\bibinfo {author} {\bibnamefont {{EHT Collaboration}}} \textit{et~al.},\ }\href {https://doi.org/10.3847/2041-8213/abe4de} {\bibfield  {journal} {\bibinfo  {journal} {The Astrophysical Journal Letters}\ }\textbf {\bibinfo {volume} {910}},\ \bibinfo {pages} {L13} (\bibinfo {year} {2021}{\natexlab{b}})}\BibitemShut {NoStop}%
\bibitem [{\citenamefont {{EHT Collaboration}}\ \emph {\textit{et~al.}}(2024)\citenamefont {{EHT Collaboration}} \emph {\textit{et~al.}}}]{EHT_SgrA_Telescope2}%
  \BibitemOpen
  \bibfield  {author} {\bibinfo {author} {\bibnamefont {{EHT Collaboration}}} \textit{et~al.},\ }\href {https://doi.org/10.3847/2041-8213/ad2df0} {\bibfield  {journal} {\bibinfo  {journal} {The Astrophysical Journal Letters}\ }\textbf {\bibinfo {volume} {964}},\ \bibinfo {pages} {L25} (\bibinfo {year} {2024})}\BibitemShut {NoStop}%
\bibitem [{\citenamefont {Narayan}\ \emph {\textit{et~al.}}(2021)\citenamefont {Narayan}, \citenamefont {Palumbo}, \citenamefont {Johnson}, \citenamefont {Gelles}, \citenamefont {Himwich}, \citenamefont {Chang}, \citenamefont {Ricarte}, \citenamefont {Dexter}, \citenamefont {Gammie}, \citenamefont {Chael} \emph {\textit{et~al.}}}]{narayan_polarized_2021}%
  \BibitemOpen
  \bibfield  {author} {\bibinfo {author} {\bibfnamefont {R.}~\bibnamefont {Narayan}}, \bibinfo {author} {\bibfnamefont {D.~C.~M.}\ \bibnamefont {Palumbo}}, \bibinfo {author} {\bibfnamefont {M.~D.}\ \bibnamefont {Johnson}}, \bibinfo {author} {\bibfnamefont {Z.}~\bibnamefont {Gelles}}, \bibinfo {author} {\bibfnamefont {E.}~\bibnamefont {Himwich}}, \bibinfo {author} {\bibfnamefont {D.~O.}\ \bibnamefont {Chang}}, \bibinfo {author} {\bibfnamefont {A.}~\bibnamefont {Ricarte}}, \bibinfo {author} {\bibfnamefont {J.}~\bibnamefont {Dexter}}, \bibinfo {author} {\bibfnamefont {C.~F.}\ \bibnamefont {Gammie}}, \bibinfo {author} {\bibfnamefont {A.~A.}\ \bibnamefont {Chael}}, \textit{et~al.},\ }\href {https://doi.org/10.3847/1538-4357/abf117} {\bibfield  {journal} {\bibinfo  {journal} {The Astrophysical Journal}\ }\textbf {\bibinfo {volume} {912}},\ \bibinfo {pages} {35} (\bibinfo {year} {2021})}\BibitemShut {NoStop}%
\bibitem [{\citenamefont {Ferrario}\ \emph {\textit{et~al.}}(2015)\citenamefont {Ferrario}, \citenamefont {Melatos},\ and\ \citenamefont {Zrake}}]{Ferrario_2015}%
  \BibitemOpen
  \bibfield  {author} {\bibinfo {author} {\bibfnamefont {L.}~\bibnamefont {Ferrario}}, \bibinfo {author} {\bibfnamefont {A.}~\bibnamefont {Melatos}},\ and\ \bibinfo {author} {\bibfnamefont {J.}~\bibnamefont {Zrake}},\ }\href {https://doi.org/10.1007/s11214-015-0138-y} {\bibfield  {journal} {\bibinfo  {journal} {Space Science Reviews}\ }\textbf {\bibinfo {volume} {191}},\ \bibinfo {pages} {77} (\bibinfo {year} {2015})}\BibitemShut {NoStop}%
\bibitem [{\citenamefont {Cremaschini}\ and\ \citenamefont {Stuchl\'{\i}k}(2013)}]{dynamo_effects_2013}%
  \BibitemOpen
  \bibfield  {author} {\bibinfo {author} {\bibfnamefont {C.}~\bibnamefont {Cremaschini}}\ and\ \bibinfo {author} {\bibfnamefont {Z.}~\bibnamefont {Stuchl\'{\i}k}},\ }\href {https://doi.org/10.1103/PhysRevE.87.043113} {\bibfield  {journal} {\bibinfo  {journal} {Phys. Rev. E}\ }\textbf {\bibinfo {volume} {87}},\ \bibinfo {pages} {043113} (\bibinfo {year} {2013})}\BibitemShut {NoStop}%
\bibitem [{\citenamefont {Latif}\ and\ \citenamefont {Schleicher}(2016)}]{Latif_2016}%
  \BibitemOpen
  \bibfield  {author} {\bibinfo {author} {\bibfnamefont {M.~A.}\ \bibnamefont {Latif}}\ and\ \bibinfo {author} {\bibfnamefont {D.~R.~G.}\ \bibnamefont {Schleicher}},\ }\href {https://doi.org/10.1051/0004-6361/201527266} {\bibfield  {journal} {\bibinfo  {journal} {Astronomy Astrophysics}\ }\textbf {\bibinfo {volume} {585}},\ \bibinfo {pages} {A151} (\bibinfo {year} {2016})}\BibitemShut {NoStop}%
\bibitem [{\citenamefont {Grasso}\ and\ \citenamefont {Rubinstein}(2001)}]{Dario_2001}%
  \BibitemOpen
  \bibfield  {author} {\bibinfo {author} {\bibfnamefont {D.}~\bibnamefont {Grasso}}\ and\ \bibinfo {author} {\bibfnamefont {H.~R.}\ \bibnamefont {Rubinstein}},\ }\href {https://doi.org/https://doi.org/10.1016/S0370-1573(00)00110-1} {\bibfield  {journal} {\bibinfo  {journal} {Physics Reports}\ }\textbf {\bibinfo {volume} {348}},\ \bibinfo {pages} {163} (\bibinfo {year} {2001})}\BibitemShut {NoStop}%
\bibitem [{\citenamefont {Jain}\ and\ \citenamefont {Sloth}(2012)}]{Jain_2012}%
  \BibitemOpen
  \bibfield  {author} {\bibinfo {author} {\bibfnamefont {R.~K.}\ \bibnamefont {Jain}}\ and\ \bibinfo {author} {\bibfnamefont {M.~S.}\ \bibnamefont {Sloth}},\ }\href {https://doi.org/10.1103/PhysRevD.86.123528} {\bibfield  {journal} {\bibinfo  {journal} {Phys. Rev. D}\ }\textbf {\bibinfo {volume} {86}},\ \bibinfo {pages} {123528} (\bibinfo {year} {2012})}\BibitemShut {NoStop}%
\bibitem [{\citenamefont {Subramanian}(2016)}]{Subramanian_2016}%
  \BibitemOpen
  \bibfield  {author} {\bibinfo {author} {\bibfnamefont {K.}~\bibnamefont {Subramanian}},\ }\href {https://doi.org/10.1088/0034-4885/79/7/076901} {\bibfield  {journal} {\bibinfo  {journal} {Reports on Progress in Physics}\ }\textbf {\bibinfo {volume} {79}},\ \bibinfo {pages} {076901} (\bibinfo {year} {2016})}\BibitemShut {NoStop}%
\bibitem [{\citenamefont {Wang}\ and\ \citenamefont {Jia}(2024)}]{Qianchuan_2023}%
  \BibitemOpen
  \bibfield  {author} {\bibinfo {author} {\bibfnamefont {Q.}~\bibnamefont {Wang}}\ and\ \bibinfo {author} {\bibfnamefont {J.}~\bibnamefont {Jia}},\ }\href {https://doi.org/10.1088/1674-1137/ad4018} {\bibfield  {journal} {\bibinfo  {journal} {Chinese Physics C}\ }\textbf {\bibinfo {volume} {48}},\ \bibinfo {pages} {085103} (\bibinfo {year} {2024})}\BibitemShut {NoStop}%
\bibitem [{\citenamefont {Tian-Xi}\ and\ \citenamefont {Yong-Jiu}(2009)}]{Gong_2009}%
  \BibitemOpen
  \bibfield  {author} {\bibinfo {author} {\bibfnamefont {G.}~\bibnamefont {Tian-Xi}}\ and\ \bibinfo {author} {\bibfnamefont {W.}~\bibnamefont {Yong-Jiu}},\ }\href {https://doi.org/10.1088/0256-307X/26/3/030402} {\bibfield  {journal} {\bibinfo  {journal} {Chinese Physics Letters}\ }\textbf {\bibinfo {volume} {26}},\ \bibinfo {pages} {030402} (\bibinfo {year} {2009})}\BibitemShut {NoStop}%
\bibitem [{\citenamefont {Boyer}\ and\ \citenamefont {Lindquist}(1967)}]{Boyer_Lindquist_Metric}%
  \BibitemOpen
  \bibfield  {author} {\bibinfo {author} {\bibfnamefont {R.~H.}\ \bibnamefont {Boyer}}\ and\ \bibinfo {author} {\bibfnamefont {R.~W.}\ \bibnamefont {Lindquist}},\ }\href {https://doi.org/10.1063/1.1705193} {\bibfield  {journal} {\bibinfo  {journal} {Journal of Mathematical Physics}\ }\textbf {\bibinfo {volume} {8}},\ \bibinfo {pages} {265} (\bibinfo {year} {1967})}\BibitemShut {NoStop}%
\bibitem [{\citenamefont {Prasanna}\ and\ \citenamefont {Varma}(1977)}]{Prasanna_1977}%
  \BibitemOpen
  \bibfield  {author} {\bibinfo {author} {\bibfnamefont {A.~R.}\ \bibnamefont {Prasanna}}\ and\ \bibinfo {author} {\bibfnamefont {R.~K.}\ \bibnamefont {Varma}},\ }\href {https://doi.org/10.1007/BF02847416} {\bibfield  {journal} {\bibinfo  {journal} {Pramana}\ }\textbf {\bibinfo {volume} {8}},\ \bibinfo {pages} {229} (\bibinfo {year} {1977})}\BibitemShut {NoStop}%
\bibitem [{\citenamefont {Prasanna}\ and\ \citenamefont {Vishveshwara}(1978)}]{Prasanna_1978}%
  \BibitemOpen
  \bibfield  {author} {\bibinfo {author} {\bibfnamefont {A.~R.}\ \bibnamefont {Prasanna}}\ and\ \bibinfo {author} {\bibfnamefont {C.~V.}\ \bibnamefont {Vishveshwara}},\ }\href {https://doi.org/10.1007/BF02848160} {\bibfield  {journal} {\bibinfo  {journal} {Pramana}\ }\textbf {\bibinfo {volume} {11}},\ \bibinfo {pages} {359} (\bibinfo {year} {1978})}\BibitemShut {NoStop}%
\bibitem [{\citenamefont {He}\ \emph {\textit{et~al.}}(2024)\citenamefont {He}, \citenamefont {Xu},\ and\ \citenamefont {Jia}}]{He_2024}%
  \BibitemOpen
  \bibfield  {author} {\bibinfo {author} {\bibfnamefont {J.}~\bibnamefont {He}}, \bibinfo {author} {\bibfnamefont {S.}~\bibnamefont {Xu}},\ and\ \bibinfo {author} {\bibfnamefont {J.}~\bibnamefont {Jia}},\ }\href {https://doi.org/10.1140/epjc/s10052-024-13680-y} {\bibfield  {journal} {\bibinfo  {journal} {The European Physical Journal C}\ }\textbf {\bibinfo {volume} {84}},\ \bibinfo {pages} {1292} (\bibinfo {year} {2024})}\BibitemShut {NoStop}%
\bibitem [{\citenamefont {Wald}(1974)}]{Wald_1974}%
  \BibitemOpen
  \bibfield  {author} {\bibinfo {author} {\bibfnamefont {R.~M.}\ \bibnamefont {Wald}},\ }\href {https://doi.org/10.1103/PhysRevD.10.1680} {\bibfield  {journal} {\bibinfo  {journal} {Phys. Rev. D}\ }\textbf {\bibinfo {volume} {10}},\ \bibinfo {pages} {1680} (\bibinfo {year} {1974})}\BibitemShut {NoStop}%
\bibitem [{\citenamefont {Aliev}\ and\ \citenamefont {Gal'tsov}(1989)}]{Aliev_Galtsov_1989}%
  \BibitemOpen
  \bibfield  {author} {\bibinfo {author} {\bibfnamefont {A.~N.}\ \bibnamefont {Aliev}}\ and\ \bibinfo {author} {\bibfnamefont {D.~V.}\ \bibnamefont {Gal'tsov}},\ }\href {https://doi.org/10.1070/PU1989v032n01ABEH002677} {\bibfield  {journal} {\bibinfo  {journal} {Soviet Physics Uspekhi}\ }\textbf {\bibinfo {volume} {32}},\ \bibinfo {pages} {75} (\bibinfo {year} {1989})}\BibitemShut {NoStop}%
\bibitem [{\citenamefont {Aliev}\ and\ \citenamefont {Özdemir}(2002)}]{Aliev_Ozdemir_2002}%
  \BibitemOpen
  \bibfield  {author} {\bibinfo {author} {\bibfnamefont {A.~N.}\ \bibnamefont {Aliev}}\ and\ \bibinfo {author} {\bibfnamefont {N.}~\bibnamefont {Özdemir}},\ }\href {https://doi.org/10.1046/j.1365-8711.2002.05727.x} {\bibfield  {journal} {\bibinfo  {journal} {Monthly Notices of the Royal Astronomical Society}\ }\textbf {\bibinfo {volume} {336}},\ \bibinfo {pages} {241} (\bibinfo {year} {2002})}\BibitemShut {NoStop}%
\bibitem [{\citenamefont {Owens}\ and\ \citenamefont {Forsyth}(2013)}]{Mathew_2013}%
  \BibitemOpen
  \bibfield  {author} {\bibinfo {author} {\bibfnamefont {M.~J.}\ \bibnamefont {Owens}}\ and\ \bibinfo {author} {\bibfnamefont {R.~J.}\ \bibnamefont {Forsyth}},\ }\href {https://doi.org/10.12942/lrsp-2013-5} {\bibfield  {journal} {\bibinfo  {journal} {Living Reviews in Solar Physics}\ }\textbf {\bibinfo {volume} {10}},\ \bibinfo {pages} {5} (\bibinfo {year} {2013})}\BibitemShut {NoStop}%
\bibitem [{\citenamefont {Hathaway}(2015)}]{HaleCycle_2015}%
  \BibitemOpen
  \bibfield  {author} {\bibinfo {author} {\bibfnamefont {D.~H.}\ \bibnamefont {Hathaway}},\ }\href {https://doi.org/10.1007/lrsp-2015-4} {\bibfield  {journal} {\bibinfo  {journal} {Living Reviews in Solar Physics}\ }\textbf {\bibinfo {volume} {12}},\ \bibinfo {pages} {4} (\bibinfo {year} {2015})}\BibitemShut {NoStop}%
\bibitem [{\citenamefont {Yang}\ \emph {\textit{et~al.}}(2024)\citenamefont {Yang}, \citenamefont {Tian}, \citenamefont {Tomczyk}, \citenamefont {Liu}, \citenamefont {Gibson}, \citenamefont {Morton},\ and\ \citenamefont {Downs}}]{Zihao_2024}%
  \BibitemOpen
  \bibfield  {author} {\bibinfo {author} {\bibfnamefont {Z.}~\bibnamefont {Yang}}, \bibinfo {author} {\bibfnamefont {H.}~\bibnamefont {Tian}}, \bibinfo {author} {\bibfnamefont {S.}~\bibnamefont {Tomczyk}}, \bibinfo {author} {\bibfnamefont {X.}~\bibnamefont {Liu}}, \bibinfo {author} {\bibfnamefont {S.}~\bibnamefont {Gibson}}, \bibinfo {author} {\bibfnamefont {R.~J.}\ \bibnamefont {Morton}},\ and\ \bibinfo {author} {\bibfnamefont {C.}~\bibnamefont {Downs}},\ }\href {https://doi.org/10.1126/science.ado2993} {\bibfield  {journal} {\bibinfo  {journal} {Science}\ }\textbf {\bibinfo {volume} {386}},\ \bibinfo {pages} {76} (\bibinfo {year} {2024})}\BibitemShut {NoStop}%
\bibitem [{\citenamefont {Wang}\ and\ \citenamefont {Sheeley}(2003)}]{Wang_2003}%
  \BibitemOpen
  \bibfield  {author} {\bibinfo {author} {\bibfnamefont {Y.-M.}\ \bibnamefont {Wang}}\ and\ \bibinfo {author} {\bibfnamefont {N.~R.}\ \bibnamefont {Sheeley}, \bibfnamefont {Jr.}},\ }\href {https://doi.org/10.1086/375026} {\bibfield  {journal} {\bibinfo  {journal} {The Astrophysical Journal}\ }\textbf {\bibinfo {volume} {590}},\ \bibinfo {pages} {1111} (\bibinfo {year} {2003})}\BibitemShut {NoStop}%
\bibitem [{\citenamefont {Wang}\ and\ \citenamefont {Sheeley}(2009)}]{Wang_2009}%
  \BibitemOpen
  \bibfield  {author} {\bibinfo {author} {\bibfnamefont {Y.-M.}\ \bibnamefont {Wang}}\ and\ \bibinfo {author} {\bibfnamefont {N.~R.}\ \bibnamefont {Sheeley}},\ }\href {https://doi.org/10.1088/0004-637X/694/1/L11} {\bibfield  {journal} {\bibinfo  {journal} {The Astrophysical Journal}\ }\textbf {\bibinfo {volume} {694}},\ \bibinfo {pages} {L11} (\bibinfo {year} {2009})}\BibitemShut {NoStop}%
\bibitem [{\citenamefont {Park}\ \emph {\textit{et~al.}}(2017)\citenamefont {Park}, \citenamefont {Folkner}, \citenamefont {Konopliv}, \citenamefont {Williams}, \citenamefont {Smith},\ and\ \citenamefont {Zuber}}]{Park_2017}%
  \BibitemOpen
  \bibfield  {author} {\bibinfo {author} {\bibfnamefont {R.~S.}\ \bibnamefont {Park}}, \bibinfo {author} {\bibfnamefont {W.~M.}\ \bibnamefont {Folkner}}, \bibinfo {author} {\bibfnamefont {A.~S.}\ \bibnamefont {Konopliv}}, \bibinfo {author} {\bibfnamefont {J.~G.}\ \bibnamefont {Williams}}, \bibinfo {author} {\bibfnamefont {D.~E.}\ \bibnamefont {Smith}},\ and\ \bibinfo {author} {\bibfnamefont {M.~T.}\ \bibnamefont {Zuber}},\ }\href {https://doi.org/10.3847/1538-3881/aa5be2} {\bibfield  {journal} {\bibinfo  {journal} {The Astronomical Journal}\ }\textbf {\bibinfo {volume} {153}},\ \bibinfo {pages} {121} (\bibinfo {year} {2017})}\BibitemShut {NoStop}%
\bibitem [{\citenamefont {Petrie}(2022)}]{Petrie_2022}%
  \BibitemOpen
  \bibfield  {author} {\bibinfo {author} {\bibfnamefont {G.~J.~D.}\ \bibnamefont {Petrie}},\ }\href {https://doi.org/10.3847/1538-4357/aca1a8} {\bibfield  {journal} {\bibinfo  {journal} {The Astrophysical Journal}\ }\textbf {\bibinfo {volume} {941}},\ \bibinfo {pages} {142} (\bibinfo {year} {2022})}\BibitemShut {NoStop}%
\bibitem [{\citenamefont {Prsa}\ \emph {\textit{et~al.}}(2016)\citenamefont {Prsa}, \citenamefont {Harmanec}, \citenamefont {Torres}, \citenamefont {Mamajek}, \citenamefont {Asplund}, \citenamefont {Capitaine}, \citenamefont {Christensen-Dalsgaard}, \citenamefont {Depagne}, \citenamefont {Haberreiter}, \citenamefont {Hekker} \emph {\textit{et~al.}}}]{Prsa_2016}%
  \BibitemOpen
  \bibfield  {author} {\bibinfo {author} {\bibfnamefont {A.}~\bibnamefont {Prsa}}, \bibinfo {author} {\bibfnamefont {P.}~\bibnamefont {Harmanec}}, \bibinfo {author} {\bibfnamefont {G.}~\bibnamefont {Torres}}, \bibinfo {author} {\bibfnamefont {E.}~\bibnamefont {Mamajek}}, \bibinfo {author} {\bibfnamefont {M.}~\bibnamefont {Asplund}}, \bibinfo {author} {\bibfnamefont {N.}~\bibnamefont {Capitaine}}, \bibinfo {author} {\bibfnamefont {J.}~\bibnamefont {Christensen-Dalsgaard}}, \bibinfo {author} {\bibfnamefont {E.}~\bibnamefont {Depagne}}, \bibinfo {author} {\bibfnamefont {M.}~\bibnamefont {Haberreiter}}, \bibinfo {author} {\bibfnamefont {S.}~\bibnamefont {Hekker}}, \textit{et~al.},\ }\href {https://doi.org/10.3847/0004-6256/152/2/41} {\bibfield  {journal} {\bibinfo  {journal} {The Astronomical Journal}\ }\textbf {\bibinfo {volume} {152}},\ \bibinfo {pages} {41} (\bibinfo {year} {2016})}\BibitemShut {NoStop}%
\bibitem [{\citenamefont {Luzum}\ \emph {\textit{et~al.}}(2011)\citenamefont {Luzum}, \citenamefont {Capitaine}, \citenamefont {Fienga}, \citenamefont {Folkner}, \citenamefont {Fukushima}, \citenamefont {Hilton}, \citenamefont {Hohenkerk}, \citenamefont {Krasinsky}, \citenamefont {Petit},\ and\ \citenamefont {Pitjeva}}]{Luzum_2011}%
  \BibitemOpen
  \bibfield  {author} {\bibinfo {author} {\bibfnamefont {B.}~\bibnamefont {Luzum}}, \bibinfo {author} {\bibfnamefont {N.}~\bibnamefont {Capitaine}}, \bibinfo {author} {\bibfnamefont {A.}~\bibnamefont {Fienga}}, \bibinfo {author} {\bibfnamefont {W.}~\bibnamefont {Folkner}}, \bibinfo {author} {\bibfnamefont {T.}~\bibnamefont {Fukushima}}, \bibinfo {author} {\bibfnamefont {J.}~\bibnamefont {Hilton}}, \bibinfo {author} {\bibfnamefont {C.}~\bibnamefont {Hohenkerk}}, \bibinfo {author} {\bibfnamefont {G.}~\bibnamefont {Krasinsky}}, \bibinfo {author} {\bibfnamefont {G.}~\bibnamefont {Petit}},\ and\ \bibinfo {author} {\bibfnamefont {E.}~\bibnamefont {Pitjeva}},\ }\href {https://doi.org/10.1007/s10569-011-9352-4} {\bibfield  {journal} {\bibinfo  {journal} {Celestial Mechanics and Dynamical Astronomy}\ }\textbf {\bibinfo {volume} {110}} (\bibinfo {year} {2011})}\BibitemShut {NoStop}%
\bibitem [{\citenamefont {Habibi}\ \emph {\textit{et~al.}}(2017)\citenamefont {Habibi}, \citenamefont {Gillessen}, \citenamefont {Martins}, \citenamefont {Eisenhauer}, \citenamefont {Plewa}, \citenamefont {Pfuhl}, \citenamefont {George}, \citenamefont {Dexter}, \citenamefont {Waisberg}, \citenamefont {Ott} \emph {\textit{et~al.}}}]{Habibi_2017}%
  \BibitemOpen
  \bibfield  {author} {\bibinfo {author} {\bibfnamefont {M.}~\bibnamefont {Habibi}}, \bibinfo {author} {\bibfnamefont {S.}~\bibnamefont {Gillessen}}, \bibinfo {author} {\bibfnamefont {F.}~\bibnamefont {Martins}}, \bibinfo {author} {\bibfnamefont {F.}~\bibnamefont {Eisenhauer}}, \bibinfo {author} {\bibfnamefont {P.~M.}\ \bibnamefont {Plewa}}, \bibinfo {author} {\bibfnamefont {O.}~\bibnamefont {Pfuhl}}, \bibinfo {author} {\bibfnamefont {E.}~\bibnamefont {George}}, \bibinfo {author} {\bibfnamefont {J.}~\bibnamefont {Dexter}}, \bibinfo {author} {\bibfnamefont {I.}~\bibnamefont {Waisberg}}, \bibinfo {author} {\bibfnamefont {T.}~\bibnamefont {Ott}}, \textit{et~al.},\ }\href {https://doi.org/10.3847/1538-4357/aa876f} {\bibfield  {journal} {\bibinfo  {journal} {The Astrophysical Journal}\ }\textbf {\bibinfo {volume} {847}},\ \bibinfo {pages} {120} (\bibinfo {year} {2017})}\BibitemShut {NoStop}%
\end{thebibliography}
%

\end{document}